\documentclass[twocolumn, twocolappendix]{aastex631}
\usepackage{amsmath}
\usepackage{bm}
\usepackage{amssymb}

\graphicspath{{./}{figures/}}

\begin{document}

\title{How Ram Pressure Drives Radial Gas Motions in the Surviving Disk}

\author[0000-0001-7011-9291]{Nina Akerman}\email{nina.akerman@studenti.unipd.it}
\affiliation{INAF - Astronomical Observatory of Padova, vicolo dell'Osservatorio 5, IT-35122 Padova, Italy}
\affiliation{Dipartimento di Fisica e Astronomia `Galileo Galilei', Università di Padova, vicolo dell'Osservatorio 3, IT-35122, Padova, Italy}

\author[0000-0002-8710-9206]{Stephanie Tonnesen}
\affiliation{Flatiron Institute, CCA, 162 5th Avenue, New York, NY 10010, USA}

\author[0000-0001-8751-8360]{Bianca Maria Poggianti}
\affiliation{INAF - Astronomical Observatory of Padova, vicolo dell'Osservatorio 5, IT-35122 Padova, Italy}

\author[0000-0001-5303-6830]{Rory Smith}
\affiliation{Departamento de Física, Universidad Técnica Federico Santa María, Vicuña Mackenna 3939, San Joaquín, Santiago de Chile}

\author[0000-0002-5655-6054]{Antonino Marasco}
\affiliation{INAF - Astronomical Observatory of Padova, vicolo dell'Osservatorio 5, IT-35122 Padova, Italy}

\begin{abstract}

Galaxy evolution can be dramatically affected by the environment, especially by the dense environment of a galaxy cluster. Recent observational studies show that massive galaxies undergoing strong ram pressure stripping (RPS) have an enhanced frequency of nuclear activity. We investigate this topic using a suite of wind-tunnel hydrodynamical simulations of a massive $M_\text{star} = 10^{11} M_\odot$ disk galaxy with 39 pc resolution and including star formation and stellar feedback. We find that RPS increases the inflow of gas to the galaxy centre regardless of the wind impact angle. This increase is driven by the mixing of interstellar and non-rotating intracluster media at all wind angles, and by increased torque on the inner disk gas, mainly from local pressure gradients when the ICM wind has an edge-on component. In turn, the strong pressure torques are driven by rising ram pressure. We estimate the black hole (BH) accretion using Bondi-Hoyle and torque models, and compare it with the mass flux in the central 140 pc region. We find that the torque model predicts much less accretion onto the BH of a RPS galaxy than the Bondi-Hoyle estimator. We argue that both models are incomplete --- the commonly used torque model does not account for torques caused by the gas distribution or local pressure gradients, while the Bondi-Hoyle estimator depends on the sound speed of the hot gas, which includes the ICM in stripped galaxies. An estimator that accounts for this missing physics is required to capture BH accretion in a RPS galaxy.

\end{abstract}

%% Keywords should appear after the \end{abstract} command. 
%% The AAS Journals now uses Unified Astronomy Thesaurus concepts:
%% https://astrothesaurus.org
%% You will be asked to selected these concepts during the submission process
%% but this old "keyword" functionality is maintained in case authors want
%% to include these concepts in their preprints.


\section{Introduction} \label{sec:intro}

Environment plays a crucial role in galaxy evolution, specifically driving the evolution of satellite galaxies from star-forming to quenched \citep{Dressler1980}. Some environmental mechanisms affect both the gaseous and stellar components of a galaxy. Among those are major and minor mergers \citep{Barnes1992}, galaxy harassment (the combined effect of several high-speed close galaxy-galaxy encounters) \citep{Moore1996}, and tidal stripping due to the gravitational influence of a cluster as a whole \citep{Byrd1990, Gnedin03}. Other mechanisms directly affect only the gaseous component, such as `strangulation' (or `starvation') \citep{Larson1980, Balogh00} and ram pressure stripping \citep[RPS,][]{GG72}. The latter is the focus of this paper. 

RPS occurs when a galaxy falls into a massive group or a cluster, and the intra- group or cluster medium (IGrM or ICM) exerts ram pressure (RP) on the interstellar medium (ISM) of the galaxy, removing it in the process. In some cases removed gas forms tails that can be as long as 100 kpc: H$\alpha$ \citep{Gavazzi01, Smith10, GASPXXIII}, X-ray \citep{SunVikhlinin05, Machacek05, Sun10, Sun22}, molecular gas \citep{OosterlooVanGorkom05, Jachym19}. See also reviews by \cite{vanGorkom04, Boselli22}. In case the tails are longer than the disk radius, the resulting structure is often referred to as a jellyfish galaxy, a term coined by \cite{Smith10}. RPS is most effective in clusters, where it rapidly quenches star formation (SF) in galaxies \citep{BoselliGavazzi06, Koopmann04} in an outside-in scenario \citep{Cortese12, Cramer2019, Owers19, GASPXXIV}, but it has also been identified in group systems \citep{GASPXII_Vulcani18, Roberts21, GASPXXXIII_Vulcani21, Kolcu22}.

With further study, it has become clear that quenching is not the only way in which RPS affects galaxies. Before removing a significant portion of gas, RP can enhance SF in the disk \citep{Vulcani18, GASPXXIV}, however this is not universally observed \citep{Boselli14, Boselli16}. Using hydrodynamical simulations \cite{Bekki14} and \cite{Steinhauser16} suggested that galaxies that fall into a cluster near-edge-on and on orbits that get closer to the cluster centre are more likely to have enhanced star formation \citep[see also][]{Roediger14, Lee20}. \cite{RuggieroLimaNeto17} showed that under RP, star formation rate (SFR) is initially increased regardless of cluster mass and whether or not a cluster has a cool core (but they do not find enhancement near the pericentre as galaxies get almost completely stripped). These simulations agree that, when in place, SF enhancement is due to RP compressing the gas. Yet, this effect is not strong enough to produce starburst galaxies. Moreover, though RPS is a gas-only interaction, it can have an effect on the stellar disk by affecting where new stars will be formed. The recent discovery by \cite{GASPXXIX} showed that RP can `unwind' galaxies, resulting in increased pitch angle of outer spiral arms. Only the youngest stars inhabit these unwinding arms, suggesting that as RP was pushing the gas, the stars were formed in situ. Their simulations show that in these galaxies hydrodynamical processes (the ICM-ISM interaction) alone are able to cause of the unwinding pattern.

A more recently discovered aspect of RPS is its connection to active galactic nuclei (AGN). Generally speaking, optical AGN tend to prefer the field environment ($\sim$5 per cent AGN fraction) to groups and clusters ($\sim$1 per cent) \citep{Dressler1985, Lopes17}, while radio-loud AGN prefer clusters \citep{Best05}. \cite{Martini07} found that the fraction of X-ray selected AGN does not depend on the environment, though \cite{Arnold09} showed the opposite. 

Using data from GAs Stripping Phenomena in galaxies project \citep[GASP,][]{GASPI} and selecting the most extremely stripped galaxies, \cite{GASPNature} found that 70 per cent of RPS galaxies (out of 7 objects in the sample) hosted a Seyfert 2 AGN. This percentage rises to 85 if they included LINERs. This incidence rate is considerably higher than in regular galaxies in any environment. The authors concluded that RP triggered the nuclear activity. Later, AGN-driven outflows \citep{GASPXIX} and AGN feedback \citep{GASPXVIII} were identified in a subset of those galaxies. \cite{Peluso22} confirmed the AGN-RPS connection on a bigger sample of 131 RPS candidates (51 of which are from GASP). In contrast, \cite{Roman-Oliveira19} in their study of 70 RPS galaxies identified only 5 AGN. All three studies note that AGN are hosted by the most massive galaxies with $M_\text{star}>10^{10}M_\odot$.

The question of the AGN-RPS connection has not yet been thoroughly studied in numerical simulations. Still, some works have noticed that surviving disk gas may contract towards the galaxy centre. For example, \cite{SchulzStruck01} and \cite{Tonnesen09} discussed how the loss of angular momentum draws gas clouds on lower orbits and, potentially, to the galaxy centre. If a galaxy's magnetic field is included and causes the gas disk to flare, RP can create oblique shocks with produce flows of gas from the outskirts to the centre \citep{Ramos-Martinez18}. Recently, \cite{Farber22} found that including cosmic rays decreases the amount of central inflow in RPS galaxies, though this inflow is still increased compared to non-RPS galaxies. Finally, using a cosmological simulation \cite{Ricarte20} showed that for massive ($M_\text{star}>10^{10}M_\odot$) galaxies RP increases accretion onto a black hole (BH) prior to quenching SF in galaxy.

The goal of this study is to test the circumstances under which RP increases inflow of gas to the galactic centre, estimate BH accretion, and identify the responsible mechanisms. We take a systematic approach to this problem in hydrodynamic simulations and gain full control of the initial conditions by simulating galaxies in isolation. Compared to the previous works, our simulations combine high resolution (up to 39 pc), star formation and stellar feedback and a varying RP wind, and although we do not model AGN accretion and feedback directly, we calculate the likely BH feeding rate in post-processing.

This paper is organised as follows. In Section \ref{sec:methods} we describe our simulations, with Section \ref{subsec:JO201} dedicated to the initial conditions. We examine the results starting with a global picture of stripping in Section \ref{sec:global}. In Section \ref{sec:centre_gas} we look at cold gas in the galaxy centre and analyse mechanisms that can drive the inflow of this gas in Section \ref{sec:processes}. We estimate BH accretion in Section \ref{sec:BH_acc}. We discuss our results in Section \ref{sec:discussion} and draw conclusions in Section \ref{sec:conclusions}.

\vfill\null

\section{Methodology} \label{sec:methods}

\subsection{Simulation set-up}

For our simulations we use the adaptive mesh refinement code Enzo \citep{Enzo}, solving the Euler equations of hydrodynamics using the piecewise parabolic method \citep{Colella&Woodward1984}. Our simulation box has 160 kpc on a side (maximum cell size of 1.25 kpc), with 5 levels of refinement allowing for the smallest cell of 39 pc. We use baryon mass ($\approx$7500 M$_\odot$) and Jeans length as our refinement criteria. Our chosen criteria resolve the entire simulated galaxy disk to the finest level.

To model RPS, rather than moving a galaxy through a cluster we fix it in the centre of the simulation box and add an ICM wind via inflow boundary conditions (and outflow on the opposite side), running `wind-tunnel' simulations. In order to study the role that the wind angle (i.e., the angle between the wind direction and the galaxy rotation axis) might play in funnelling gas into the galactic centre, we model three wind angles: $0^\circ$ (a face-on wind that flows along the z-direction in our simulated box, W0), $45^\circ$ (W45, in which the wind has equal components along the y- and z-directions), and $90^\circ$ (edge-on wind that flows along the y-direction in our simulated box, W90). The wind angle is fixed throughout each simulation. We also simulate a fiducial isolated galaxy that is not subject to RP (NW) as a control for comparison.

For data analysis and visualisation we use the yt python package \citep{yt}.

\subsection{Galaxy JO201} \label{subsec:JO201}

Galaxy parameter space and the parameter space of the ICM in clusters are vast. We base the initial conditions of our simulation suite on an observed galaxy that is both subject to RP and has an AGN. Galaxy JO201 is one of the most thoroughly-studied RPS galaxies \citep{GASPII,GASPXV,GASPXXVI, GASPXXXIV} and is also among the GASP AGN sample \citep{GASPNature, GASPXVIII, GASPXIX}. Modelling our simulated galaxy and wind after JO201 gives us an initial expectation that the central BH must be growing, and will allow us to directly compare to a variety of existing multi-wavelength observations in future work. We stress that our goal is not to provide a detailed numerical model for J0201, but to simply select initial conditions based on observational data.

In our simulations, we follow \cite{Tonnesen09}. We model a Plummer–Kuzmin stellar disk \citep{Miyamoto&Nagai1975} and Burkert profile for the spherical dark matter (DM) halo \citep{Burkert1995, Mori&Burkert00} as static potentials \citep[see also][]{Roediger06}, while the self-gravity of the gas component is calculated at each time step.

In order to ensure the correspondence of the initial conditions to JO201, we match the kinematics of the simulated galaxy with the real one. For that we fit the JO201 stellar rotation curve from \cite{GASPII} (their Figure 16) with our model using the aforementioned potentials. 

If we restrict the mass of the stellar disk to $M_\text{star} = 10^{11} M_\odot$, the resulting best-fit curve has the following parameters: stellar disk mass $M_\text{star} = 10^{11} M_\odot$, scale length $r_\text{star} = 5.94$ kpc and scale height $z_\text{star} = 0.58$ kpc, and the core radius of the DM halo $r_\text{DM} = 17.36$ kpc. We also compare the parameters to the observational data to make sure that they are within the observational error margins \citep{GASPII,GASPXV,GASPXXVI}. Note that we exclude the bulge from this simulation because we lack measurements of this structure in JO201. The gaseous disk mass is 10 per cent of the stellar disk mass $M_\text{gas} = 10^{10} M_\odot$ and its scale length is $r_\text{gas} = 1.7 \times r_\text{star} = 10.1$ kpc.

The last missing ingredients in the initial conditions are the wind and the pre-wind ICM parameters. \cite{Tonnesen19} showed that modelling a wind varying in strength (instead of constant) is essential to understand the evolution of the gas disk in RPS galaxies. In order to model a realistic infall of JO201 into its cluster we follow the procedure described in \cite{GASPXV}, where the shape of the potential well of the cluster is derived assuming hydrostatic equilibrium of an isothermal ICM with temperature 7.1 $\pm$ 0.2 keV and a beta profile matching the observed X-ray distribution \citep{GASPXXXIV}. The galaxy begins with a clustercentric radius of 1.9 Mpc and velocity of $1785 \; \mathrm {km \, s}^{-1}$. Then a time integrator is used to follow the galaxy acceleration as it falls into the cluster, resulting in the evolution of the galaxy velocity (which translates to ICM velocity in our `wind-tunnel' simulations) and ICM density. We consider this a good estimation for the orbit of JO201 given the uncertainties arising from projection effects. The simplifying assumption that the RP increases with time, while the wind angle stays constant allows us to both model a realistic RP profile and to study the effects of the wind direction on a galaxy evolution. The ICM temperature is assumed to be constant $T=7.55\times10^7$ K. We calculate the pre-wind ICM conditions from the initial ICM wind parameters, using Rankine–Hugoniot jump conditions assuming Mach number of 3 as in \cite{Tonnesen09, Tonnesen19}.

\vfill\null

\subsection{Included physics}

Our simulations use the chemistry and cooling library {\sc grackle}\footnote{\url{https://grackle.readthedocs.io}} \citep{Smith17}. {\sc grackle} includes a primordial cooling routine for atomic H, He, H$_2$ and HD, as well as cooling due to metal line emission and photoheating rates, with the cooling floor of 10 K, and a Haardt \& Madau UV background \citep{Haard&Madaut12}. The disk starts with metallicity $Z = 1.0 \; Z_\odot$ and the ICM has $Z = 0.3 \; Z_\odot$.

For SF and stellar feedback recipes we follow \cite{Goldbaum15, Goldbaum16}. The main parameter of this SF model is the minimum threshold number density $n_{\rm{thresh}}$ at which SF can happen in a cell, if its mass exceeds the Jeans mass. We assume $n_{\rm{thresh}}=10 \, \rm{cm}^{-3}$, minimum star particle mass is $10^3 M_\odot$, and SF efficiency is 1 per cent. For a galaxy in isolation, such parameters result in a SFR of 2 $M_\odot$yr$^{-1}$ that matches the observational value for local galaxies of stellar mass $M_\text{star} = 10^{11} M_\odot$ \citep{BrinchmannEllis00, Zheng07, Pannella09, Bauer13}. We also analysed the SF history of isolated galaxies with $n_{\rm{thresh}} = 3 \, \rm{cm}^{-3}$ and $30 \, \rm{cm}^{-3}$ and found them to evolve similarly, meaning that the parameter choice will not significantly influence our results. Moreover, on the timescale of $\sim100$ Myr, the SFR surface density converges to match the global Kennicutt-Schmidt relation, for any of the three $n_{\rm{thresh}}$ values. Stellar feedback includes ionising radiation from young stars (heating up to $10^4$ K), winds from evolved massive stars and the energy and momentum released by individual supernova (SN) explosions (combined energy budget of $10^{51}$ erg, which first adds the terminal momentum input from the number of SN expected during a given time step, then adds any additional energy in the form of thermal energy). SN also increase the metallicity of a cell. \cite{Goldbaum16} shows that this stellar feedback model reproduces the observed local Kennicutt-Schmidt relation. Their resolution is 20 pc, somewhat higher than ours of 39 pc, therefore we confirm that with our chosen $n_{\rm{thresh}}$ we also find agreement with the observed local Kennicutt-Schmidt law.

The galaxy is first evolved for 230 Myr in isolation. During this time the disk settles down, as the variation of SFR on a 5 Myr timescale goes down from 300 per cent to 5 per cent. At that point we start the wind inflow through the lower boundary of the simulation box and restart our simulation as four separate runs: our three wind runs and the continued isolated NW galaxy. An additional 70 Myr is needed for the wind to reach the galaxy disc, so that the wind reaches a galaxy at 300 Myr.

Note that our simulations do not include a BH seed, a sink particle that can consume gas in simulations. Instead, we will estimate BH accretion in post-processing using several accretion models.

\section{Global picture} \label{sec:global}

\begin{figure*}
\centering
\includegraphics[width = 150mm]{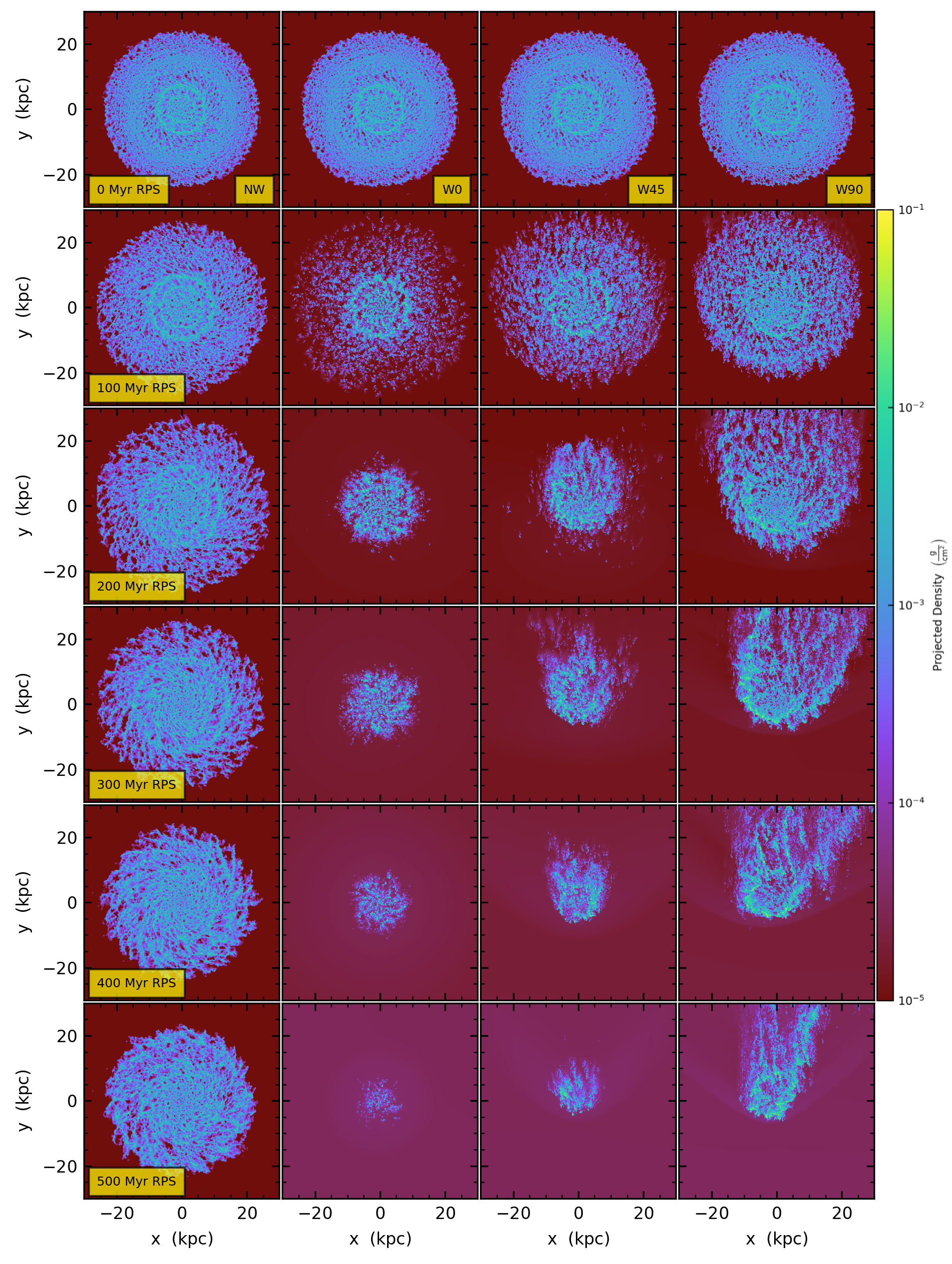}
\caption{Density projections of the central 30 kpc of the simulated region, within $\pm5$ kpc of the disk plane. For RPS-galaxies (W0, W45 and W90), the process of stripping can be followed, while the fiducial (NW) galaxy does not evolve significantly.}
\label{Fig:dens_proj_30}
\end{figure*}

\begin{figure}[t]
\includegraphics[width = 85mm]{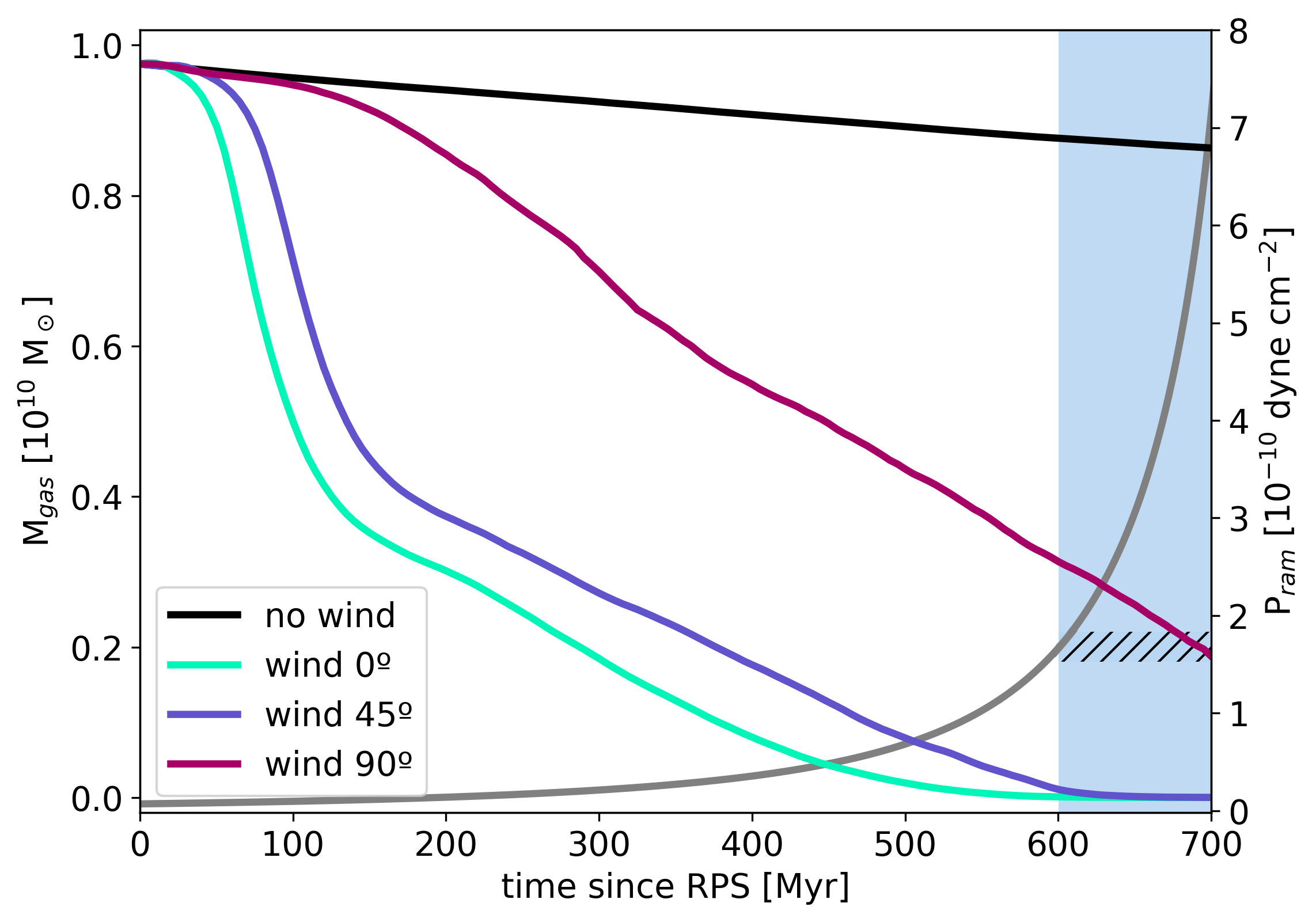}
\caption{Total gas mass with metallicity $Z>0.7 \, Z_\odot$ within a galaxy disk (defined with a radius 30 kpc and a height $\pm5$ kpc from the disk plane) as a function of time and RP, colour-coded by wind angle. Ram pressure $P_\text{ram}$ as a function of time as grey line. Blue area indicates the expected RP JO201 is currently experiencing based on its position and velocity and on the density of its parent cluster \citep{GASPIX, ErratumGASPIX}, and the hatched area shows its approximate gas mass.}
\label{Fig:tot_mass}
\end{figure}

\begin{figure}[t]
\includegraphics[width = 85mm]{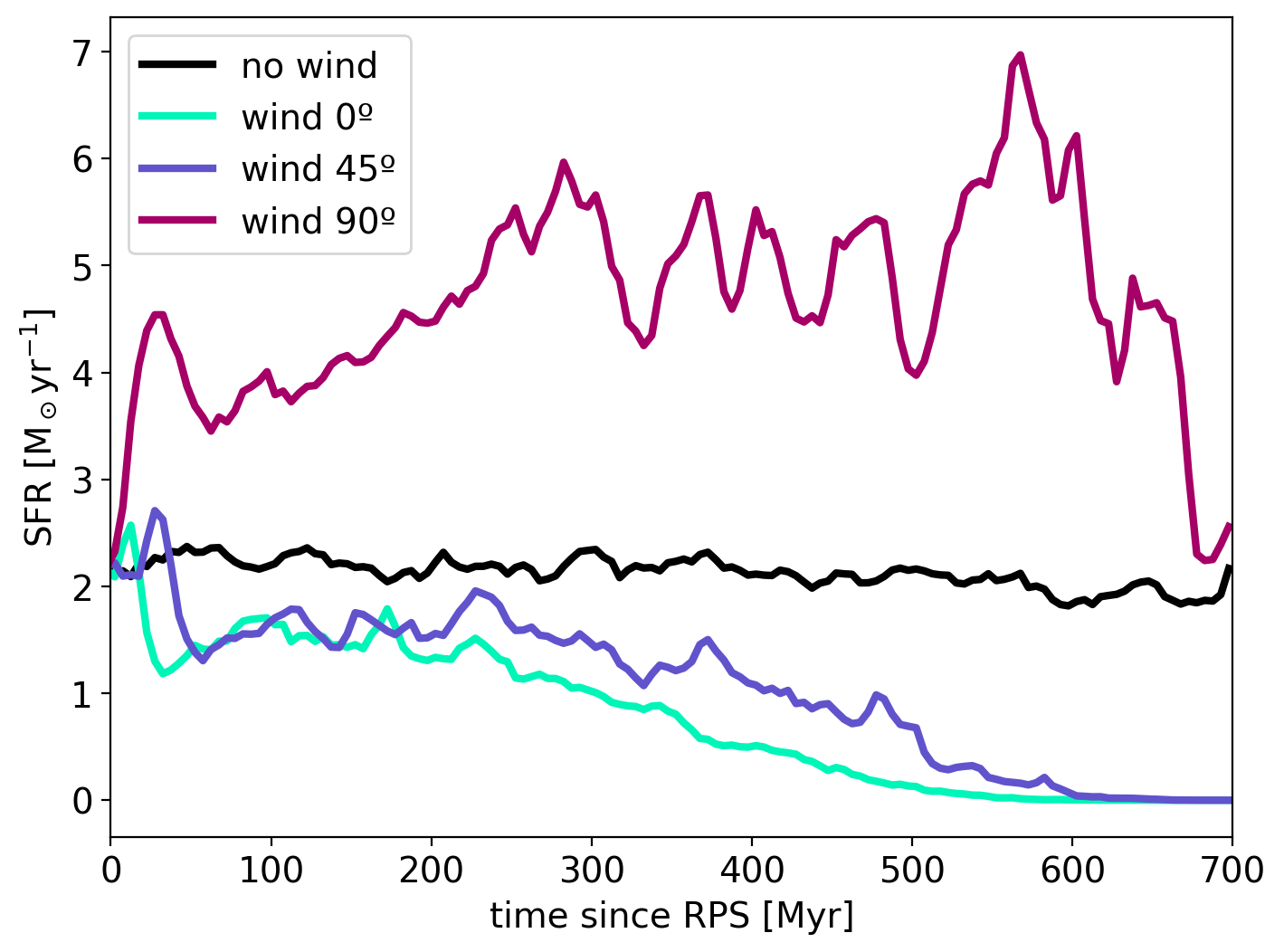}
\caption{Star formation history within a galaxy disk (defined with a radius 30 kpc and a height $\pm5$ kpc from the disk plane), colour-coded by wind angle. As with the gas mass evolution, W0 and W45 show similar SFRs while the W90 SFR is dramatically different.}
\label{Fig:SFH}
\end{figure}

To allow for straightforward comparisons with previous RPS simulations, we first examine how global galaxy properties evolve under RP.

To gain intuition on the four simulations discussed in this work, in Figure \ref{Fig:dens_proj_30} we show a time series of density projections of all gas within $\pm5$ kpc of the disk plane. Note that in Figure \ref{Fig:dens_proj_30} and every other figure the time labels (and later x-axis) are `time since RPS'. Thus, $t = 0$ marks the start of RPS and does not include the first 300 Myr during which galaxies evolve with a lower density surrounding medium and no wind. At 0 Myr, the galaxies all look nearly identical. This is by construction, as all the simulations start from the same seed and differ only when the wind enters the box. We note that at the earlier times (0 Myr and 100 Myr), a ring feature is seen in all the simulations at a radius of about 10 kpc. As discussed in \citet{Goldbaum15} \citep[see also][]{Behrendt15}, these rings appear early due to gravitational instability, then fragment. By the 0 Myr output the galaxy has been evolving for 300 Myr, so any central rings (within 5 kpc) have already fragmented, even though the 10 kpc ring is still visible. Moreover, holes induced due to initial collapse and fragmentation of gas can be seen throughout the NW evolution. Importantly, this Figure allows us to track the process of stripping as it happens in galaxies that are hit by the wind under different angles. The stripping proceeds outside-in \citep[as expected][]{Quilis2000, Kronberger08, Fumagalli14}, and by 500 Myr there is very little gas left in W0 and W45. W90 retains more gas because the leading edge of the disk shields the rest of the galaxy, with the stripped disk being asymmetric due to the combined effects of RP and galaxy rotation. Though we expect W45 to be a `case in-between' of W0 and W90, it is much closer to the former while showing some asymmetry in the gas distribution.

We quantify these differences by analysing the evolution of gas mass (Figure \ref{Fig:tot_mass}). To measure the gas mass of a galaxy we select a disk region around it with radius 30 kpc and height $\pm$5 kpc from the disk plane. We only select the gas with metallicity $Z>0.7 \, Z_\odot$. This allows us to exclude the pure ICM, while capturing some of the mixing of ISM and ICM. We note that most of this gas, $\gtrsim$ 90 per cent, is cold (T $\le$ 10$^{4.5}$ K, see Section \ref{sec:centre_gas}). We also plot the evolution of RP as a grey line. The blue area indicates the approximate RP that JO201 is currently subject to based on its position and velocity and on the density of its parent cluster \citep{GASPIX, ErratumGASPIX}.

The NW galaxy slowly consumes its gas due to SF. W0 and W45 evolve similarly to one another, with steady removal in the first 150 Myr, followed by slower, continuous stripping. The delay between W0 and W45 is because the angled wind has a larger distance to travel to reach the galaxy at the centre of the simulation box, rather than a physical delay in the stripping rate. W90, on the other hand, does not go through any initial rapid gas removal. Hence, as in \cite{Roediger06} we find that W0 and W45 show similar stripping rates and W90 shows slower gas removal.

Figure \ref{Fig:SFH} shows the evolution of the star formation rates (SFR) of our simulated galaxies. In the NW galaxy SFR shows little to no evolution over the course of the simulation, as discussed above. Like with the gas mass change, here W0 and W45 evolve similarly, experiencing a brief increase in SFR when the wind has only just reached the galaxies, followed by rapid depression of SF. The SF in these galaxies continues to decrease, and never recovers to the pre-wind SFR. W90, on the other hand, shows a SF enhancement of 2-3 times the SFR of the NW galaxy ($\sim90$ per cent of this enhancement takes place within the inner 10 kpc of the galactic disk, indicating that the enhancement is not due to SF in the stripped tail). Only when the W90 galaxy has lost 80 per cent of its gas mass does the SFR precipitously drop to the isolated galaxy SFR.

\section{Centremost cold gas} \label{sec:centre_gas}

\begin{figure*}[t]
\centering
\includegraphics[width = 150mm]{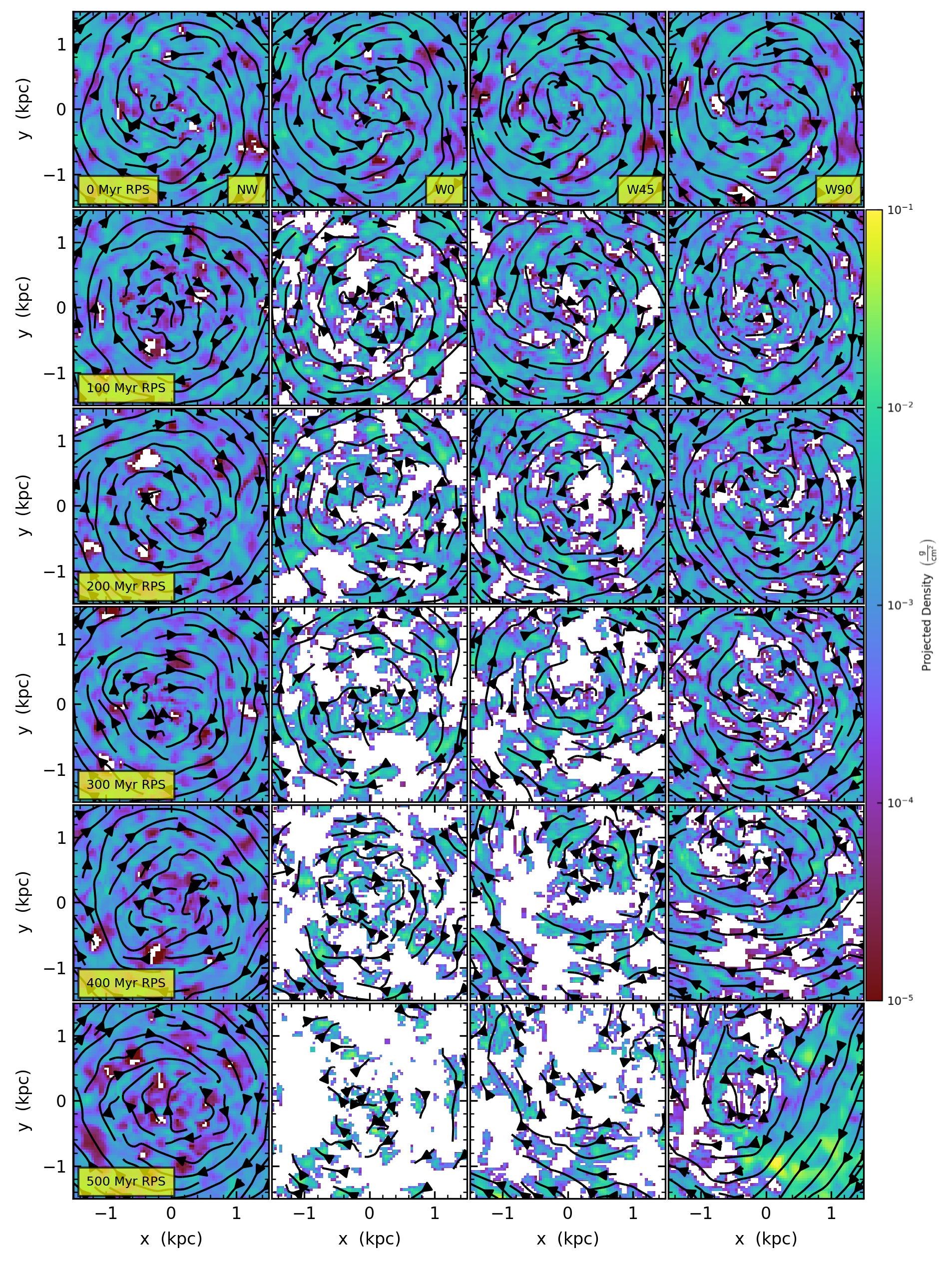}
\caption{Density projections of cold gas (T $\le$ 10$^{4.5}$ K) in the central 1.5 kpc of the disk, within $\pm500$ pc of the disk plane with overlaid velocity streamlines. The no-wind (NW) galaxy shows little evolution with time, while the wind runs show an increasing amount of hot gas which is the ICM passing through the galaxy. In this central region, W90 is the most similar to NW until late times. The circular motion of gas is clearly visible pre-RPS and throughout the evolution of NW, and is retained even after RPS substantially disturbs the gas distribution (400 Myr W0, 300 Myr W45, 500 Myr W90).}
\label{Fig:dens_proj_centre}
\end{figure*}

We now move closer to the galaxy centre, stepping towards the region where a BH would dominate the potential. We first highlight that the early gas loss in the RPS galaxies is largely in the outskirts, as illustrated by the global density projections in Figure \ref{Fig:dens_proj_30}. In Figure \ref{Fig:dens_proj_centre} we show snapshots of column density of the cold (T $\le$ 10$^{4.5}$ K) gas within $\pm500$ pc of the disk plane in the central 1.5 kpc of the disk with overlaid planar velocity streamlines. The temperature cut for the cold gas is set in order to easily compare with previous works by, e.g., \cite{Angles-Alcazar21}, who posit that this is the gas phase that can efficiently feed a BH. Although we want to include the disk gas only, as this is the gas that is most likely to accrete onto a BH, we do not require an additional a metallicity cut since the temperature cut already selects only the high metallicity gas $Z>0.7 \, Z_\odot$. 

Velocity streamlines in Fig. \ref{Fig:dens_proj_centre} highlight the circular motion of the gas, visible during all the evolutionary stages of the NW galaxy. Already at 100 Myr the central regions show different density distributions, with the wind runs (especially W0) showing more blank `holes' that, if not for the temperature cut, would be filled with high-temperature ICM. As time passes, more holes are visible in all the wind runs, but the circular motion is retained even after RPS substantially disturbs the gas distribution (400 Myr W0, 300 Myr W45, 500 Myr W90). Through comparison with a radiative-cooling only run (to be discussed in Akerman et al., in prep), we find that the holes can also be seen even in the absence of stellar feedback, which suggests they are not feedback holes that are later grown by RP \citep[holes are also seen in the radiative cooling-only runs in][]{Tonnesen09}. Note that even though the holes dominate the central gas projection in W0 and W45 as early as 300 Myr, there is still more than 20 per cent of the original gas mass left in the galaxy, and based on the projections in Figure \ref{Fig:dens_proj_30}, one would argue that the truncation radius is 5 or 10 kpc in W45 and W0, respectively. We verified that we are indeed seeing holes through the disk rather than large-scale vertical motions of the remaining gas disk. By closely examining the region near the disk plane (not shown here), we found that the impact of the ICM can induce minor vertical sloshing of the disk. However, the motion is always within 400 pc, so the disk would always be seen in these projections.

In both of the angled wind runs, W45 and W90, we also see that the point around which gas rotates shifts from the centre of the galaxy, located at the origin in Figure \ref{Fig:dens_proj_centre}, which is the potential minimum. It shifts towards the upper right in the panels, sometimes with additional whirls formed due to disordered motions in the disk, as the wind pushes the gas along the y-axis. Importantly, we expect the BH to remain at the galaxy centre (x = y = 0 in this projection) because the static potentials from the stellar disk and DM halo dramatically dominate over the gas mass. Interestingly, although in W45 this shift occurs at about the same time as in W90, the effect is overshadowed by the large holes from the ICM streaming through low density regions of the disk. In W90, on the other hand, the shift in W90 does not result in large gaps in the central region within the first 500 Myr. This is because gas is only moving within the disk plane, so gas from the disk edges is being pushed into the galaxy centre rather than above the galaxy. This is most evident at the 500 Myr projection, where an `arm' of dense gas can be seen about to sweep through the galaxy centre (this arm can be more globally seen in Figure \ref{Fig:dens_proj_30}).

Prevalence of the ICM-filled holes in W0 and W45, asymmetric mass distribution, and mis-centred gas motions in W45 and W90 all influence the amount of gas that could reach a BH. Starting with this visual frame of reference, in the rest of this section we will quantify these differences by measuring gas and stellar masses in the centre as well as mass fluxes. We will also refer to this picture throughout the paper.

\begin{figure}[t]
\centering
\includegraphics[width = 85mm]{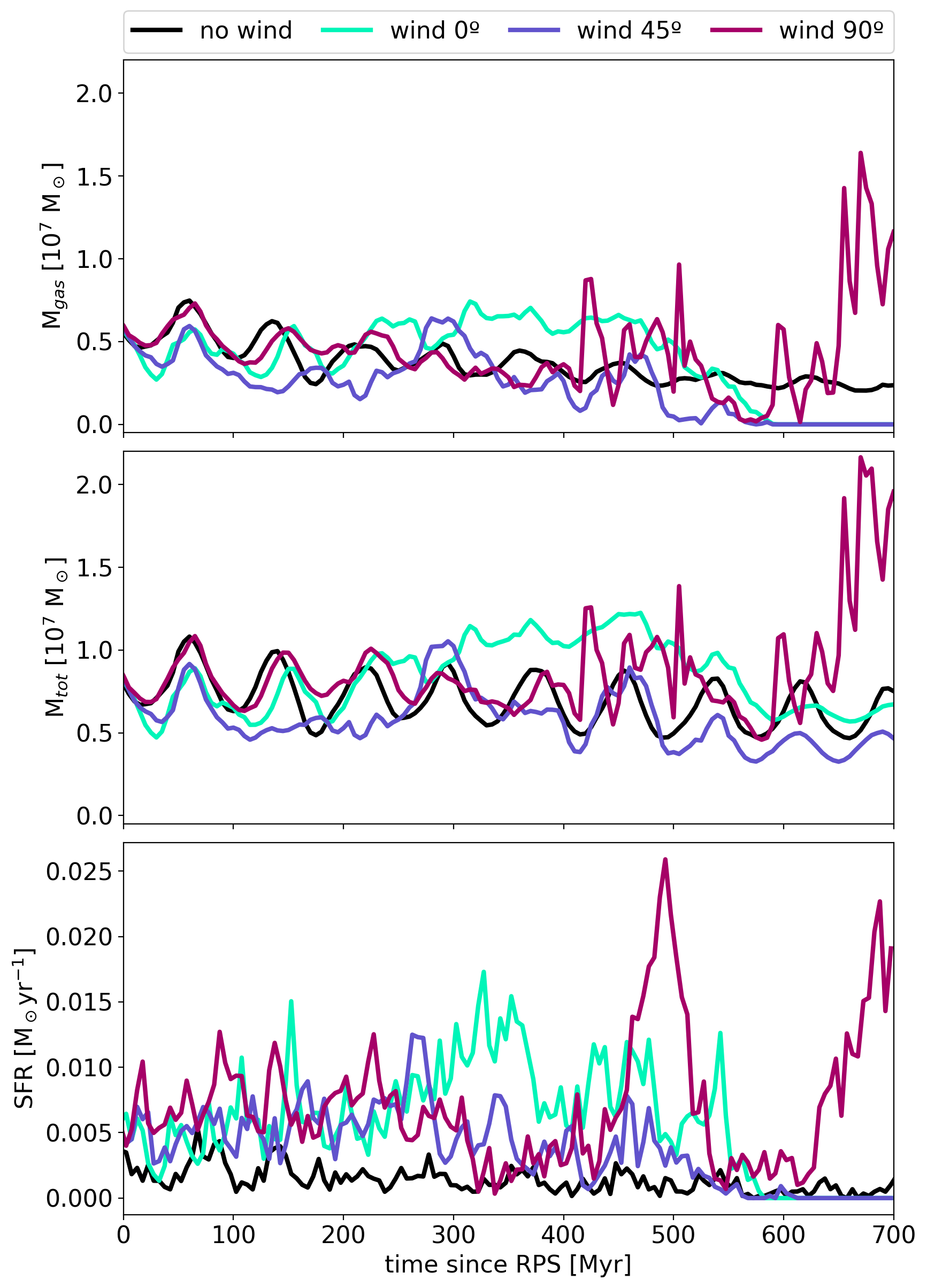}
\caption{From top to bottom: cold gas mass, total (gas with metallicity $Z>0.7 \, Z_\odot$ and stars) mass and SFR within a 500 pc sphere centred on the galaxy centre as a function of time, colour-coded by wind angle.}
\label{Fig:mass_500}
\end{figure}

In Figure \ref{Fig:mass_500} we begin to quantify the differences in the central regions of these galaxies. In the top panel we plot the cold gas mass within a galactocentric sphere of $R = 500$ pc as a function of time. We choose a 500 pc sphere because it is relatively well-resolved in all of our galaxies, with more than 10 cells across the radius of the region. We also verified that using a radius of 1 kpc does not qualitatively change our results. In NW, cold gas mass contributes from 70 (at t=0 Myr) to 30 (at t=700 Myr) per cent of the total (gas with metallicity $Z>0.7 \, Z_\odot$ plus stars) mass in 500 pc, with this total mass remaining relatively constant (i.e. the reduction in cold gas mass is due to transformation into stars, Figure \ref{Fig:mass_500} middle). The cold gas also constitutes $\geq$97 per cent of the gas with metallicity $Z>0.7 \, Z_\odot$ (and drops to 80 per cent only after 500 Myr in RPS-galaxies). Due to SF, the absolute values of cold gas mass decrease with time, and in order to remove this systematic decrease that occurs in all simulations, we compare NW and RPS galaxies at the same time step to focus on the effect of RPS.

We note that the oscillations in mass in the central region (especially well seen in NW) are mainly driven by the initial collapse of gas in the disk from radiative cooling. Star formation feedback damps the oscillations over time (we confirm this in the simulation). In Appendix \ref{appendix:delayed} we show that starting the wind at later times when these oscillations are smaller does not qualitatively affect our results.

When we focus on the central region of the galaxy we find a dramatically different picture comparing the gas mass of RPS galaxies versus the NW galaxy. First, at early times the gas mass is very similar between all four galaxies ($\le$ 200 Myr). This may be expected based on a picture of outside-in gas stripping due to RP. However, after 200 - 300 Myr we start seeing differences between the simulations, with the RPS galaxies tending to have more cold gas in their central 500 pc.

In more detail, we see more cold gas in the central region of W0 starting at about 300 Myr, lasting until about 500 Myr when the disk is nearly completely stripped. Surprisingly based on the global gas mass evolution, W0 and W45 differ in their central regions, with W45 tracking NW much more closely and losing central gas somewhat earlier than W0. W90 also shows different behaviour in the central region, tracking the NW galaxy closely until, starting at 400 Myr and lasting through the end of the simulation, W90 generally has more cold gas in the centre.

Unlike the global SFR trends shown in Fig. \ref{Fig:SFH}, SFR in the central region (Figure \ref{Fig:mass_500} bottom panel) is enhanced under RP regardless of the wind angle, with the peaks of SFR delayed with respect to the peaks in cold gas mass in the top panel of Figure \ref{Fig:mass_500}. However, as seen by comparing the top and middle panels of Figure \ref{Fig:mass_500}, M$_\text{tot}$ --- including both gas and stars --- is dominated by the cold gas mass throughout W90 and until late times in W0 and W45 (after 400 Myr) when the central gas starts to be removed. Therefore, in the central region, both the stellar and cold gas mass is enhanced due to RP.

\begin{figure*}[t]
\centering
\includegraphics[width = 160mm]{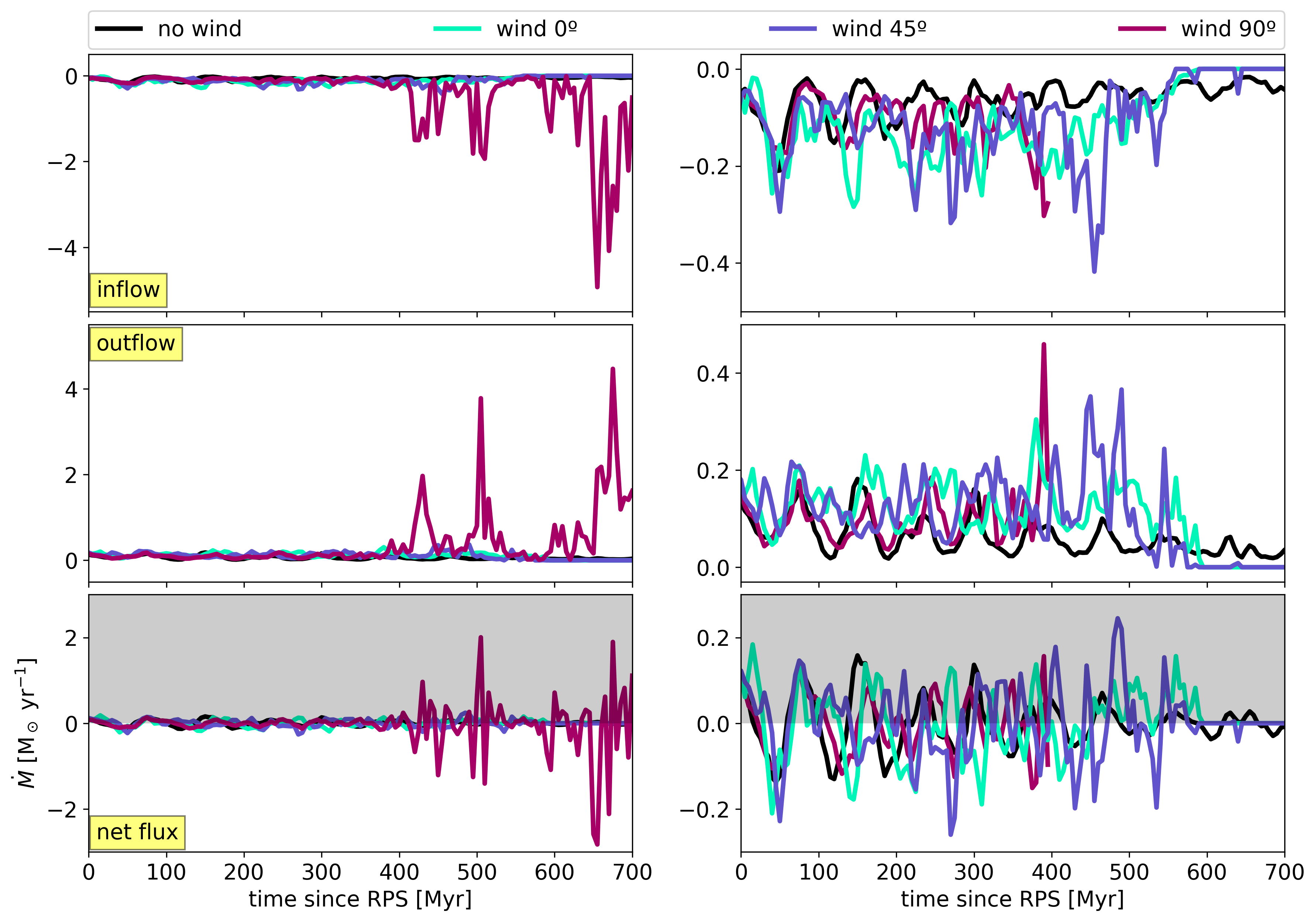}
\caption{Cold gas mass flux through a a spherical shell centred at the galaxy centre, with a radius of 500 pc and thickness of 200 pc.} From top to bottom the rows show inflow, outflow, and net flux, colour-coded by wind angle. Left and right panels show the same data, but the right panel is scaled down and W90 is shown only for the first 400 Myr of RPS. The grey area in the bottom panels serves to guide the eye and separate the inflow (negative values on white) from the outflow (positive values on grey).
\label{Fig:mass_flux}
\end{figure*}

Comparing Figures \ref{Fig:dens_proj_centre} and \ref{Fig:mass_500} also highlights another important point when considering the gas in the centre of the galaxy --- although the T $\leq10^{4.5}$ K gas in NW is more smoothly distributed than in the wind runs, the total mass in the central region is often higher in galaxies undergoing RPS. The densest clumps contain most of the mass.

Now, to quantify how the gas moves in and out of the region we calculate the mass fluxes. We select cells of cold gas in a spherical shell of radii $400 \, \text{pc} < R < 600 \, \text{pc}$, so that the shell is centred at 500 pc, consistent with Figure \ref{Fig:mass_500}. Mass flux within the shell is a sum of mass fluxes of individual cells:

\begin{equation}
 \dot{M} = \sum_i \frac{m_i v_i}{dL},
 \label{eq:mass_flux}
\end{equation}

where $m_i$ is mass of the $i$-th cell, $v_i$ is the radial (spherical $r$) velocity and $dL=200$ pc is the shell width. Here, we separate $v_i$ into positive and negative values to find outflow and inflow, respectively. Net flux is then sum of outflow and inflow.

Figure \ref{Fig:mass_flux} shows inflows, outflows and net fluxes for different wind angles and the NW case. Note that the two columns show the same data, but in the right panels W90 is shown only for the first 400 Myr of RPS and the y-axis range is scaled down to better show W0 and W45. The grey area in the bottom panels serves to guide the eye and separate the inflow from the outflow. 

Under RP, both inflows (top) and outflows (middle) of cold gas are increased. It is true for all of the wind angles, but the effect is stronger for W90 where the increase is more than an order of magnitude, which can be seen by comparing the y-axis ranges of the left and the right panels. This is also evident in Figure \ref{Fig:dens_proj_centre} as at 400 and 500 Myr the gas does not just spiral to the centre, but travels right through it because of the off-centre rotation. We have verified by integrating the net mass fluxes that they predict more gas within the central 500 pc of the galaxies affected by RP, in agreement with Figure \ref{Fig:mass_500}.

We want to highlight two important points. First, that independently of the wind angle, the radial flow of gas in the centre of the disk increases under RP. Second, as with the total mass in the central region, what is especially intriguing is that even though we are looking only at the centremost 500 pc, the increase in mass flux happens almost immediately after the wind hits the W0 and W45 galaxies. In the net flux (bottom panels), this increase in inflows and outflows is reflected as an increase in the peak values, with the difference between the wind runs and NW increasing with time until gas is completely removed due to RP. As discussed above, any vertical sloshing of the disk is within 400 pc so will not have an impact on these flux measurements. Net outflows under RP are, of course, no surprise, although we would expect this to happen later when RP is stronger. In fact, we can make an analytical estimate of when the central 500 pc should be stripped using the \cite{GG72} estimate, and find that central stripping should occur when RP is above $\sim$ 5 $\times$ 10$^{-11}$ dyne cm$^{-2}$, or 450 Myr into our simulations. Instead, we emphasise here that during RPS, the inflow of gas to the galactic centre is increased as well, sometimes by orders of magnitude in short bursts, and now we will focus on the mechanisms that can drive this inflow. 

\section{Mechanisms that drive central gas motions} \label{sec:processes}

So far we have seen that under RP both the central gas mass and the net fluxes of gas are increased. Here, we will discuss some of the possible mechanisms that could drive gas towards the galaxy centre.

\subsection{Fallback of stripped gas} \label{subsec:fallback}

\begin{figure}[t]
\includegraphics[width = 85mm]{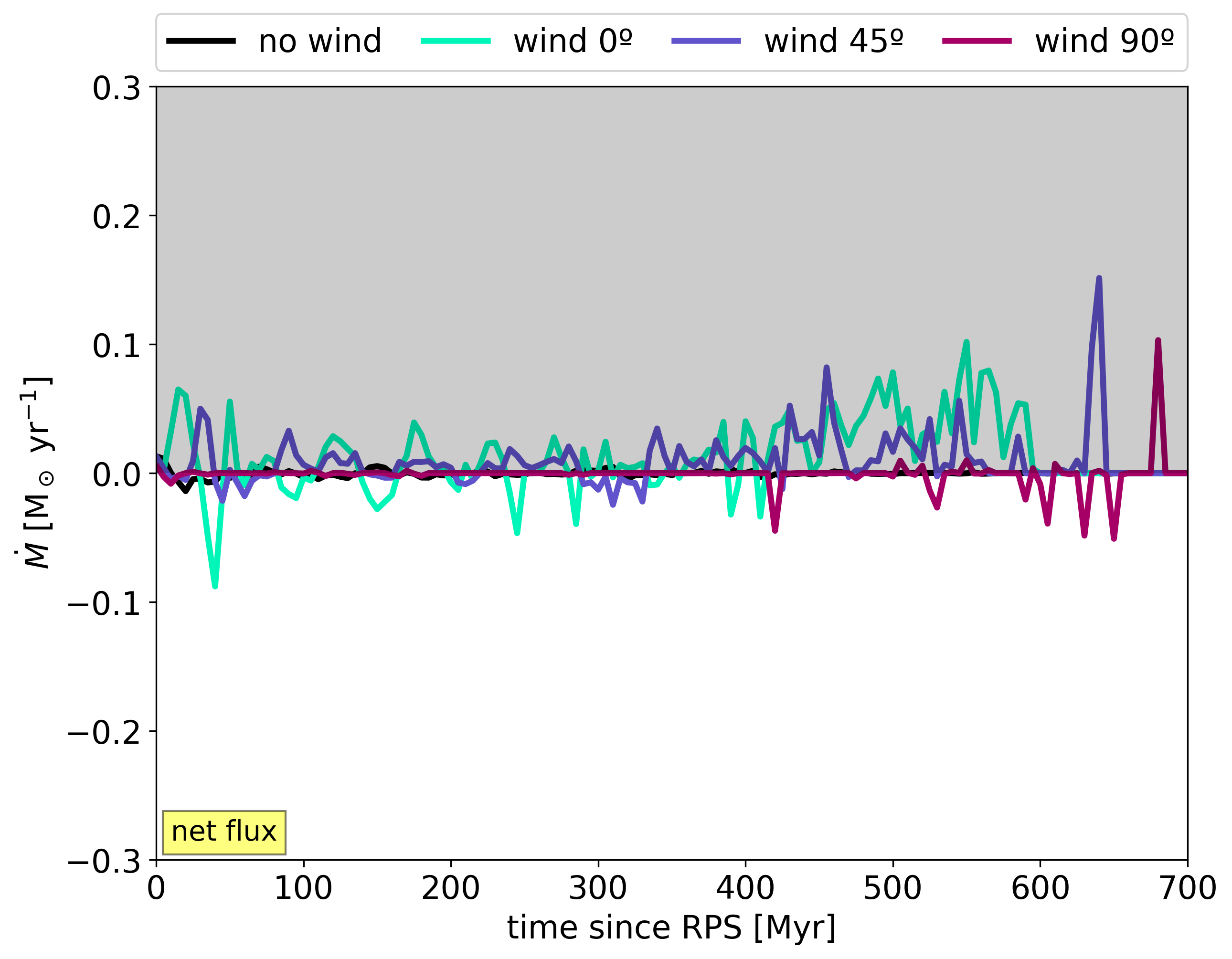}
\caption{Vertical net flux of cold gas through a cylinder centred at 500 pc height from the galaxy plane (see text), colour-coded by wind angle. The grey area serves to guide the eye and separate the inflow (negative values on white) from the outflow (positive values on grey). The y-axis range is the same as in Figure \ref{Fig:mass_flux} (right) and illustrates that vertical movement of cold gas is much smaller than mass fluxes from radial velocity.}
\label{Fig:vz_mass_flux}
\end{figure}

Fallback occurs when stripped gas that is still gravitationally bound falls back onto the galaxy \citep{Vollmer01, Tonnesen09}. To determine if fallback is responsible for the increased inflows to the galactic centre, we measure mass fluxes as in eq. \ref{eq:mass_flux}, where instead $v_i$ is vertical velocity $v_z$. Because the disk is in the x-y plane, using $v_z$ allows us to select the velocity component that is moving into or out of the plane of the disk. Since a spherical shell is not an appropriate choice for this measurement, we select a cylindrical region that is as close as possible to our 500-pc shell in Section \ref{sec:centre_gas}. First, we separate the galaxy into downwind ($z>0$, the side where the tail is in W0 and W45) and upwind ($z<0$) halves. In each half, we define a cylinder of radius $R=500$ pc and thickness of $dL=200$ pc (the same as the spherical shell's width). We place this cylinder 500 pc above (and below) the disk plane and centred on the galaxy rotation axis, and measure the flux through each cylinder. To avoid confusion in the upwind half, where negative $v_z$ are directed away from the galaxy, indicating outflow (and vice versa), we here change the sign, so that negative values always mean inflow to the galaxy centre.

Figure \ref{Fig:vz_mass_flux} shows the combined net flux of cold gas in both the galaxy halves. Compared to radial mass fluxes in the bottom panels of Figure \ref{Fig:mass_flux}, the vertical flux is several times smaller, meaning that the bulk of the gas moves in the galactic plane even under RP. Moreover, though vertical inflows are increased in RPS-galaxies, the net flux is still outflow-dominated.

We conclude that fallback does not account for the increased inflows of cold gas to the galactic centre.

\subsection{Mixing of ISM with ICM} \label{subsec:mixing}

We saw in Figure \ref{Fig:dens_proj_centre} big holes in the galaxy centre, their number and size increasing with time. These holes are filled with ICM, so we would expect some part of it to be mixing with the ISM. Since the ISM is rotation-supported gas, while the ICM is not, the mixed gas would lose angular momentum and fall to the centre \citep[as posited in][]{Tonnesen12}.

To quantify the degree of mixing using the gas metallicity as a proxy, we measure the ICM fraction in each cell:

\begin{equation}
f_{\text{ICM}} = \frac{Z_\text{cell} - Z_\text{max}}{Z_\text{ICM} - Z_\text{max}},
\end{equation}

where $Z_\text{cell}$ is metallicity of a cell, $Z_\text{ICM}=0.3 Z_\odot$ is the metallicity of the ICM, and $Z_\text{max}$ is the maximum metallicity of a star particle in the central 500 pc sphere of each galaxy at each point in time (although the size of the region has little effect as long as it includes star particles). By selecting the maximum metallicity we account for stellar feedback that continuously dumps metals into the galaxy, and we choose stellar metallicity because it is more reliable for our simulations as the gas is continuously mixed to lower metallicity and removed from the galaxy. 

\begin{figure}[t]
\includegraphics[width = 85mm]{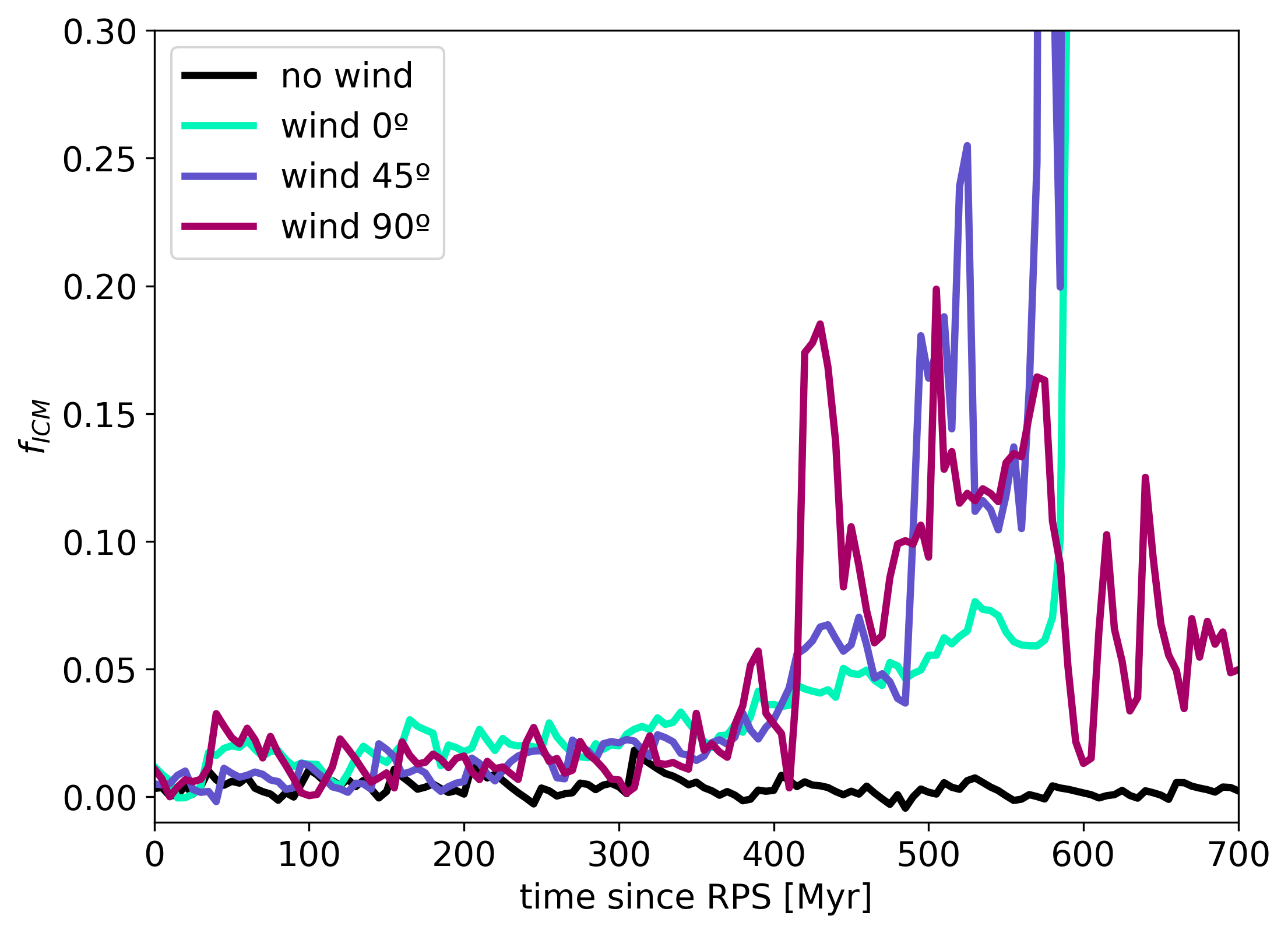}
\caption{Average ICM fraction in a 500 pc sphere at the galaxy centre as a function of time, colour-coded by wind angle. The W45 and W0 lines become vertical to an ICM fraction of one once all their gas is lost, so we limit our y-range to better see earlier mixing.}
\label{Fig:icmf_500}
\end{figure}

Figure \ref{Fig:icmf_500} shows the average ICM fraction in a $R=500$ pc central sphere. The NW galaxy has a constant near-zero fraction of ICM, which is also supported by the lack of `holes' in Figure \ref{Fig:dens_proj_centre}. We notice gradual mixing happening in all RPS-galaxies (also illustrated in Figure \ref{Fig:dens_proj_centre} with the growing number of low-temperature holes), until W0 and W45 are rapidly and completely stripped at around 600 Myr, at which point the only gas left in that region is the ICM ($f_{\text{ICM}}=1$).

It is worth examining the ICM fraction in detail in the W90 simulation. It shows very little mixing until 400 Myr, when the ICM fraction quickly increases to $\sim$0.15 for the next 200 Myr. This coincides with the increase in cold gas mass in Figure \ref{Fig:mass_500} and the increase in mass fluxes in Figure \ref{Fig:mass_flux}, which indicates that it is the cooled mixed-in ICM that is falling to the centre of the galaxy. We note that the same process is happening in all of the three RPS galaxies, but at varying levels.

During the 400-600 Myr period, the ICM fraction increases for all RPS galaxies ($\sim$0.05 for W0, 0.05-0.2 for W45 and W90). This mixed-in gas directly adds a small amount of mass into the central region that can be read from Figure \ref{Fig:icmf_500}. This adds about 5 per cent in W0, and, on average, 10 per cent in W45 and W90 of mass compared to the NW galaxy. 

In addition to this direct result of mixing, loss of angular momentum due to mixing can also add mass from larger radii within the disk to the central region. Since the ICM is non-rotating, mixing with the rotating ISM would make the mixed gas lose angular momentum and fall to a lower orbit. Therefore, some ISM gas that originates outside 500 pc will fall to within 500 pc through this process. To estimate how much gas would fall into the 500 pc sphere via angular momentum loss, we assume an average ICM fraction of 10 per cent. Because of the conservation of momentum, mixed gas becomes 10 per cent slower, which means that prior to mixing it must have been 10 per cent faster. Therefore, to find gas that will fall to 500 pc due to mixing-induced angular momentum loss, we merely need to find the radius at which gas rotates 10 per cent faster than the gas at 500 pc. If we check the gas rotation curve of the NW galaxy (which has not mixed, see Figure \ref{Fig:icmf_500}), we find such gas at $\sim$540 pc. Hence, all of the gas 500 pc$\leq r \le$540 pc will end up within 500 pc because of mixing-induced angular momentum loss. 

To see how much of the increase in central mass in the wind-impacted galaxies this actually accounts for, we compare the ISM mass from  500--540 pc in the NW galaxy to the difference in the central masses between RPS-galaxies and the NW in Figure \ref{Fig:mass_500}. There is $6 \times 10^5$ M$_{\odot}$ from 540--500 pc in NW, so, depending on the wind angle, this would account for 10 to 20 per cent of added mass, and if we include the mass of directly mixed-in ICM, the total gas mass added to the central 500 pc region due to mixing would be roughly 15 to 30 per cent above the gas mass in NW. This can account for the extra gas we see in W45 around 300 Myr and for some of the increase in W0, but not for a significant amount of the jump in W90 after 600 Myr, which means that ICM mixing can be only partially responsible for the increased gas mass in most of our wind runs.

\subsection{Torques} \label{subsec:torques}

\begin{figure*}[t]
\centering
\includegraphics[width = 160mm]{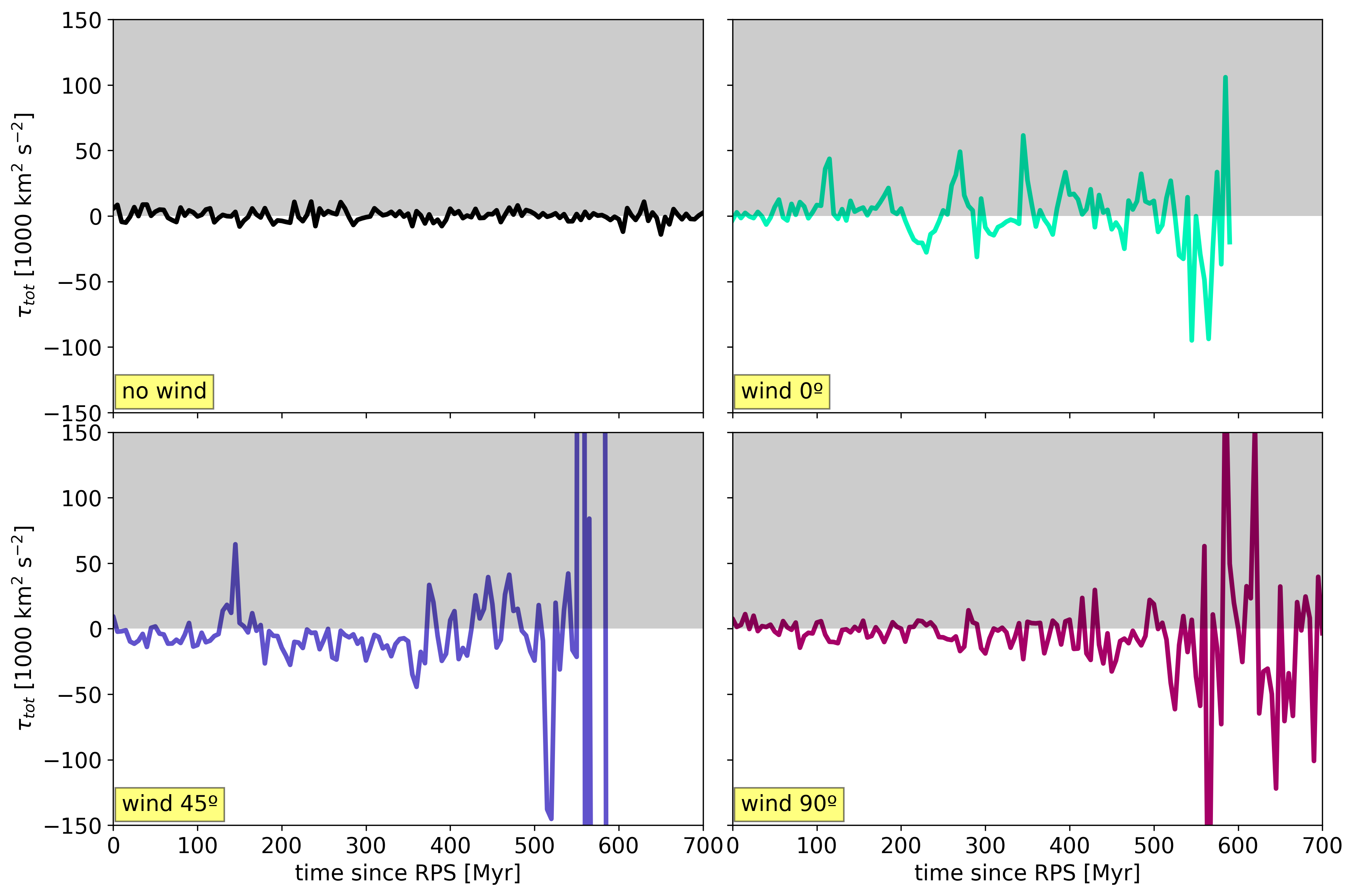}
\caption{Total (gas, stars and pressure) torque as a function of time, colour-coded by wind angle. As in Figure \ref{Fig:mass_flux}, the grey area in the bottom panels serves to guide the eye and separate torques that drive inflow (negative) from torques that drive outflow (positive).}
\label{Fig:tot_torque}
\end{figure*}

\begin{figure}
\centering
\includegraphics[width = 85mm]{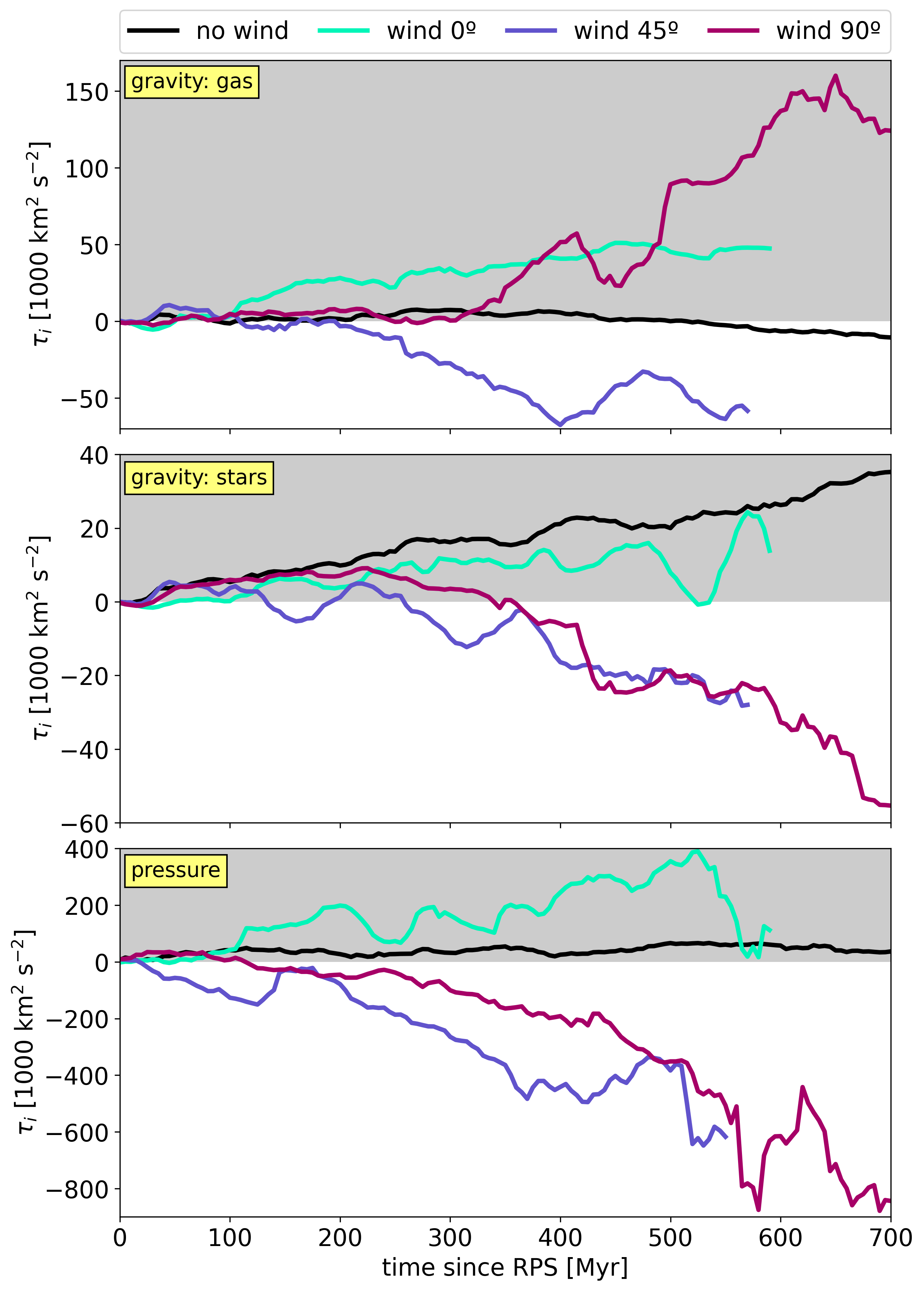}
\caption{Cumulative distribution of torques as a function of time, colour-coded by the wind angle. From top to bottom the rows show torques from gas gravity, torques from gravity of stars, and torques from pressure gradients. As in Figure \ref{Fig:mass_flux}, the grey area in the panels serves to guide the eye and separate torques that drive inflow (negative) from torques that drive outflow (positive).}
\label{Fig:torque_cumu}
\end{figure}

Another mechanism that can drive mass fluxes of gas to the galaxy centre is torque, which will change the angular momentum of gas. Torque can either be due to gravitational acceleration from an asymmetrical distribution of gas or stars (in our case the dark matter potential is spherically symmetric, therefore the torque is zero), or due to local non-radial pressure gradients. To calculate torques from the gravity of gas and stars and from pressure gradients, we use the same spherical shell at 500 pc in which we found gas fluxes in Section \ref{sec:centre_gas}, and as with mass fluxes we focus on the torque of cold gas (T $\le$ 10$^{4.5}$ K). Specific torques (per unit mass) are calculated as:

\begin{equation}
 {\bm \tau}_j = \frac{1}{M_\text{shell}} \sum_i m_i \mathbf{r}_i \times \mathbf{a}_j(\mathbf{r}_i),
\end{equation}

where $M_\text{shell} = \sum_i m_i$ is the total gas mass in the shell, $m_i$ is the mass of $i$-th cell in the shell, $\mathbf{r}_i$ is its radius-vector and $\mathbf{a}_j(\mathbf{r}_i)$ is the acceleration acting on the cell. Acceleration from different components is calculated separately. Gravitational acceleration from gas\footnote{Gravitational acceleration is found in post processing using \href{https://github.com/mikegrudic/pytreegrav}{{\sc pytreegrav}} python package by \cite{pytreegrav}.} is simply $\mathbf{a}_1 = - G \sum_i \sum_k^{i \neq k} m_k \mathbf{r}_{ik}/ r_{ik}^3$, where $m_k$ is the mass of the $k$-th cell that gravitationally acts on the $i$-th cell (these cells are from the whole simulation box), $\mathbf{r}_{ik}$ is the radius-vector between the two cells and $G$ is the gravitational constant. Gravitational acceleration $\mathbf{a}_2$ from stars is calculated using the same method, but we loop over stellar particles instead of gas cells. Finally, acceleration from local gas thermal pressure gradients is $\mathbf{a}_3 = -\nabla P(\mathbf{r}_i) / \rho(\mathbf{r}_i)$, where $P$ and $\rho$ are pressure and density, respectively. As we focus on in-plane flows, we consider only the component of the total (gas, stars and pressure) torque vector in the direction of the angular momentum vector ${\bm L}$ of the shell $\tau_\text{tot} = [{\bm \tau}_\text{gas} + {\bm \tau}_\text{stars} + {\bm \tau}_\text{press}]_{\hat{L}}$ \citep[we follow][]{Angles-Alcazar21}.

In Figure \ref{Fig:tot_torque} we plot the total (gas, stars and pressure) torque $\tau_\text{tot}$ where, as before, negative values correspond to gas inflow. As expected from the increase in measured mass fluxes (Figure \ref{Fig:mass_flux}), when we compare the total torques, the values are bigger for RPS-galaxies than for the NW. Furthermore, the periodical changes in torque qualitatively follow the periodical changes in mass fluxes, with larger net torques driving both inflows and outflows in the RPS galaxies. Upon careful examination, one can see that torques in the RPS galaxies tend to be negative and therefore drive gas inflow in W45 and W90, while the most symmetric wind, W0, shows a more even distribution of negative and positive torque values.

To gain physical understanding of what process is driving these inflows in the RPS galaxies, we can consider each torque component separately. In addition, for ease of comparison, rather than show the instantaneous torque, which fluctuates throughout the simulations, we show the cumulative torque as a function of time. In Figure \ref{Fig:torque_cumu} we show these comparisons, colour-coded by the wind angle. From top to bottom the rows show torques from gas gravity, torques from gravity of stars, and torques from pressure gradients. In the NW case, torques are dominated by pressure torques, and magnitudes of stellar and gas torques are several times smaller (as can be seen from the different y-axis ranges). Plotting cumulative distribution (as oppose to the actual values) helps us see whether each particular torque tends to drive more inflows (negative values) or outflows (positive), while also showing how they grow in time. As such, in NW pressure and stellar torques drive outflows, and gas torques change the sign to drive inflows at the end of the simulation. However, we expect that torques in general in the NW run should be minor, therefore for an intuitive interpretation use the cumulative values as a baseline magnitude below which torque should have little effect.

The relative contributions from different torques are similar in NW and wind galaxies, but under RP, magnitude of pressure and gas torques increases compared to NW, while stellar torques retain magnitudes comparable to those of the NW galaxy. We note that stellar torque in W45 and W90 tends to drive inflow, but not with strengths above NW. This agrees with our expectations that stellar torques should be the least affected as RP does not directly impact stars \citep{GG72}.

Unlike the stellar torques, the gas torques grow much more strongly with time in the RPS galaxies than in the NW galaxy, but here W0 and W90 drive outflows, and W45 drives inflow. Interestingly, the evolution of gas torque in W45 almost exactly mirrors that of W90, just with a different sign, while the cumulative torque increases much more smoothly in W0. We see that the slope of the cumulative gas torque steepens dramatically around 400 and 500 Myr, where we can also see evidence of density waves perturbing the gas at the centre of the disk in Figure \ref{Fig:dens_proj_centre}. Net torque from the gas distribution is much smaller than from local pressure gradients, but still larger than from the stellar distribution.

The growth rate of the pressure torques is similar in all RP runs and is more than an order of magnitude stronger than NW by the end of the simulations. The major difference between the wind runs is that in W0 the pressure torques tend to be positive, driving gas outflow, while for W45 and W90 the pressure torques tend to be negative, driving gas inflow.

To better understand why local pressure gradients would result in a net inflow in W45 and W90, we examined the cold gas mass distribution in our disk between 400-600 pc. We find that there is more gas with orbits going against the wind direction (since gas that orbits in the same direction as the wind is more easily stripped). Because the wind is constantly increasing, there is pressure gradient along the wind direction. This drains angular momentum from the gas orbiting against the wind and results in net inflow of gas to the galaxy centre. Similarly, \cite{Ricarte20} found that accretion onto the BH grew with increasing RP, especially at the pericentre of the orbit. Combined with our results, this indicates that inflow of gas to the galaxy centre and to the BH increases not after some RP threshold value, but because of the rising RP gradient.

We note that in order for torque to drive any real inflow in our galaxies, it must be strong enough to change the angular momentum of the gas. As torque is defined as the rate of change of angular momentum, one way to estimate the significance of torques in driving gas towards the galaxy centre is to find the timescale on which the torque would alter angular momentum. To do this we calculate the ratio of angular momentum to torque, both measured in the 500 pc shell. While there is significant variability in all of the runs (as expected from the variability of the torque seen in Figure \ref{Fig:tot_torque}), we find that the median values indicate that the torque is strong enough to impact the angular momentum on short timescales. The median timescale on which pressure torques act is 1-5 Myr, quicker for RPS galaxies as under RP the torque is growing. The stellar and the gas torques are slower, with median timescales of 10-40 Myr, since in our simulations these torques are lower than those from the pressure gradients.

We conclude that gas and pressure gradient torques grow under RP, while the stellar ones remain constant but change the sign compared to NW. In the runs with an asymmetric wind, the pressure gradient along the wind direction drives the pressure torques, and, in turn, they drive the gas to the galaxy centre. Pressure torques are larger than those from the stellar and gas distribution and, hence, dominate the net flux. However, the net flux does not seem to perfectly reflect the total torque (compare with Figure \ref{Fig:mass_flux} bottom). The most dramatic example of the mismatch is the net flux increase in W90 during 400-500 Myr that is not reflected by the torque (Figure \ref{Fig:tot_torque} bottom left). We also find, importantly, that the total net torque on W0 would result in net outflow. Therefore, gravitational and pressure gradient torques alone are not causing the gas inflow towards the galaxy centre. The other mechanism acting to drive inflows under RP, as discussed in Section \ref{subsec:mixing}, is mixing of ISM with ICM. This mixing occurs in all RPS galaxies and is likely to add 15 to 30 per cent of gas mass to their central regions depending on the wind angle.

\section{BH accretion estimation} \label{sec:BH_acc}

Now that we have determined that RPS can increase the central gas mass and examined how that happens in the different wind runs, in this section we will try to estimate BH accretion itself. We remind the reader that in our simulations we do not have a BH seed, i.e., a sink particle that could accrete the surrounding gas. Therefore, for simplicity we assume that the BH is always at the galaxy centre, which in our case is the centre of the static stellar and DM potentials, and has a mass of $10^7 M_\odot$ \citep[based on the stellar mass of $M_\text{star} = 10^{11} M_\odot$][]{BaronMenard19, vanSon19}. To estimate the BH accretion rate under the unusual conditions provided by RP we test three models: Bondi-Hoyle, torque and normalised mass fluxes. We have shown in Sections \ref{subsec:mixing} and \ref{subsec:torques} that RPS galaxies are subject to physical processes unique only to them. These processes are not accounted for by the classical models, hence we predict that they might not suffice in measuring the BH accretion.

\subsection{BH accretion models}

We first briefly introduce the three ways we estimate the accretion rate onto a central BH, comparing two commonly used BH accretion estimators to mass flux measured near the BH.

\subsubsection{Torque model}

We start by following \cite{HopkinsQuataert11}, who model accretion of cold gas that is losing angular momentum due to the presence of shocks from stellar torques. The accretion rate is:

\begin{multline}
\dot{M}_\text{BH} = \alpha f_\text{d}^{5/2} M_\text{BH,8}^{1/6} M_\text{d,9} R_{0,100}^{-3/2} (1 + f_0/f_\text{gas})^{-1} \frac{M_\odot}{\text{yr}};\\
f_0 = 0.31 f_\text{d}^2 M_\text{d,9}^{-1/3},
\label{eq:BH_acc}
\end{multline}

where $\alpha = 2.5$ is normalisation factor, $M_\text{BH,8} = M_\text{BH}/10^8 M_\odot$ is BH mass, $R_{0,100} = R_0/100$ pc is radius of a sphere within which we estimate the following values: $M_\text{d,9} = M_\text{d}/10^9 M_\odot$ is mass of the disk (stars and gas) component, $f_d$ is disk mass fraction and $f_\text{gas} = M_\text{gas}/M_\text{d}$ is mass fraction of gas in the disk. Our disk mass fraction is $f_d = 1$, since we do not include a bulge in our galaxy model and all of the gas mass is within the galaxy disk. $R_0$ should be as small as possible, because we are trying to catch gas that is gravitationally bound to BH, ideally $R_0 < 100$ pc \citep{Angles-Alcazar17}. However, due to low resolution, we choose $R_0 = 140$ pc, an aperture that \citet{Angles-Alcazar17} have shown to give adequate results, and that contains 3 cells across the radius in our simulated galaxies.

\subsubsection{Bondi-Hoyle}

We also follow the classical Bondi-Hoyle \citep{BondiHoyle} model that describes spherically-symmetrical accretion of hot, pressure supported homogeneous gas, without including gas self-gravity. While these conditions are hardly applicable to our simulated galaxies, the Bondi-Hoyle model is frequently adopted in the literature, thus it is still very useful to infer its predicted accretion rate and compare it with that of other models. Here, to take advantage of the high resolution of our simulation and to account for rotational angular momentum support (which in cold gas dominates over pressure support) we follow the approach of \cite{Tremmel17}. The accretion rate is:

\begin{multline}
\dot{M}_\text{BH} = \alpha \frac{\pi (G M_\text{BH})^2 \rho c_s}{(v_\theta^2 + c_s^2)^2}; \\
\alpha =
    \begin{cases}
      (\frac{n}{ n_{\rm{thresh}} } )^2 & \text{if } n \geq n_{\rm{thresh}}\\
      1 & \text{if } n < n_{\rm{thresh}}
    \end{cases}
\label{eq:Bondi}
\end{multline}

Ambient density $\rho$ and speed of sound $c_s$ are measured in a sphere of radius $R_\text{B-H}=700$ pc, centred at the galaxy centre. Cold, angular momentum-supported gas that we aim to capture predominantly moves in the plane of the galaxy (Section \ref{subsec:fallback}), while the Bondi-Hoyle model requires us to include all the gas in our region, not just the cold component. Hence, we measure average angular velocity $v_\theta$ in a thin cylindrical shell of radii $660 \, \text{pc} < R < 740 \, \text{pc}$ and height $h=50$ pc. The density-dependent boost factor $\alpha$ accounts for unresolved ISM, as we compare the local number density $n$ (measured in the same cylindrical shell) to the SF density threshold $n_{\rm{thresh}}$ (Section \ref{sec:methods}). Note that the bulk velocity (typically present in Bondi-Hoyle) is $v_\text{bulk} = 0$, since we assume an unmoving BH at the galaxy centre.

Although we have the necessary resolution to get closer to the BH, we will show that selecting gas within $R_\text{B-H}=700$ pc results in the same accretion rate as calculated by the torque model and mass flux measurement in the fiducial (NW) galaxy. Since the torque model is more sophisticated and describes physical conditions that are closer to the ones found at the galaxy centre, we can assume that it more correctly models accretion when RP is not present (NW). Indeed, as we will discuss in Section \ref{subsec:BHflux}, \cite{Angles-Alcazar21} find that the torque model well-traces the mass flux close to the BH. Hence, we use the $R_\text{B-H}$ that gives us a similar accretion rate using Bondi-Hoyle. Moreover, even on these scales we stay within the Bondi radius ($2 G M_\text{BH} / c_s^2 \sim 10^3$ pc) throughout the NW simulation, although the Bondi radius shrinks in the RPS galaxies over time as the inner region is dominated by hot ICM gas. If we use a lower $R_\text{B-H}$ to find the values in equation \ref{eq:BH_acc}, we overestimate the accretion rate relative to torque and mass flux models, because of the high densities on those scales.

\subsubsection{Normalised mass flux}\label{subsec:BHflux}

For the third approach we use mass fluxes. If we measure net flux of cold gas around 1 kpc and around 140 pc (as is eq. \ref{eq:mass_flux}), in NW about 10 per cent of the gas would make it to the inner region. A similar decrease in the mass flux is seen in \cite{Angles-Alcazar21}, and this trend continues as they look closer to the BH using a hyper-resolved simulation. Based on the \cite{Angles-Alcazar21} results, we assume that this can be applied to gas moving towards the BH, and that at any point of time the fraction of gas that inflows from 140 pc to the BH stays constant. We choose a spherical shell of radius $100 \, \text{pc}< R < 180 \, \text{pc}$ that is centred at 140 pc so that comparison with the torque model is straightforward. To estimate accretion rates, we find net flux in the region; if at any point of time net flux is negative (inflow), we multiply it by a factor $\beta$; if it is outflow, we consider accretion to be zero:

\begin{multline} 
\dot{M}_\text{BH} = - \beta \dot{M}_\text{140};\;
\dot{M}_\text{140} =
    \begin{cases}
      \dot{M}_\text{140} & \text{if } \dot{M}_\text{140} < 0\\
      0 & \text{if } \dot{M}_\text{140} > 0
    \end{cases}
\label{eq:flux_acc}
\end{multline}

Here, we choose $\beta = 0.025$ so that accretion rates measured by mass fluxes and the torque model would be similar. Importantly, this also falls within the range seen when comparing the mass flux between 140 pc and the BH in the super-resolved simulations of \cite{Angles-Alcazar21} (their Figure 10).

\subsection{Comparison}\label{BH:comparison}

\begin{figure}
\centering
\includegraphics[width = 85mm]{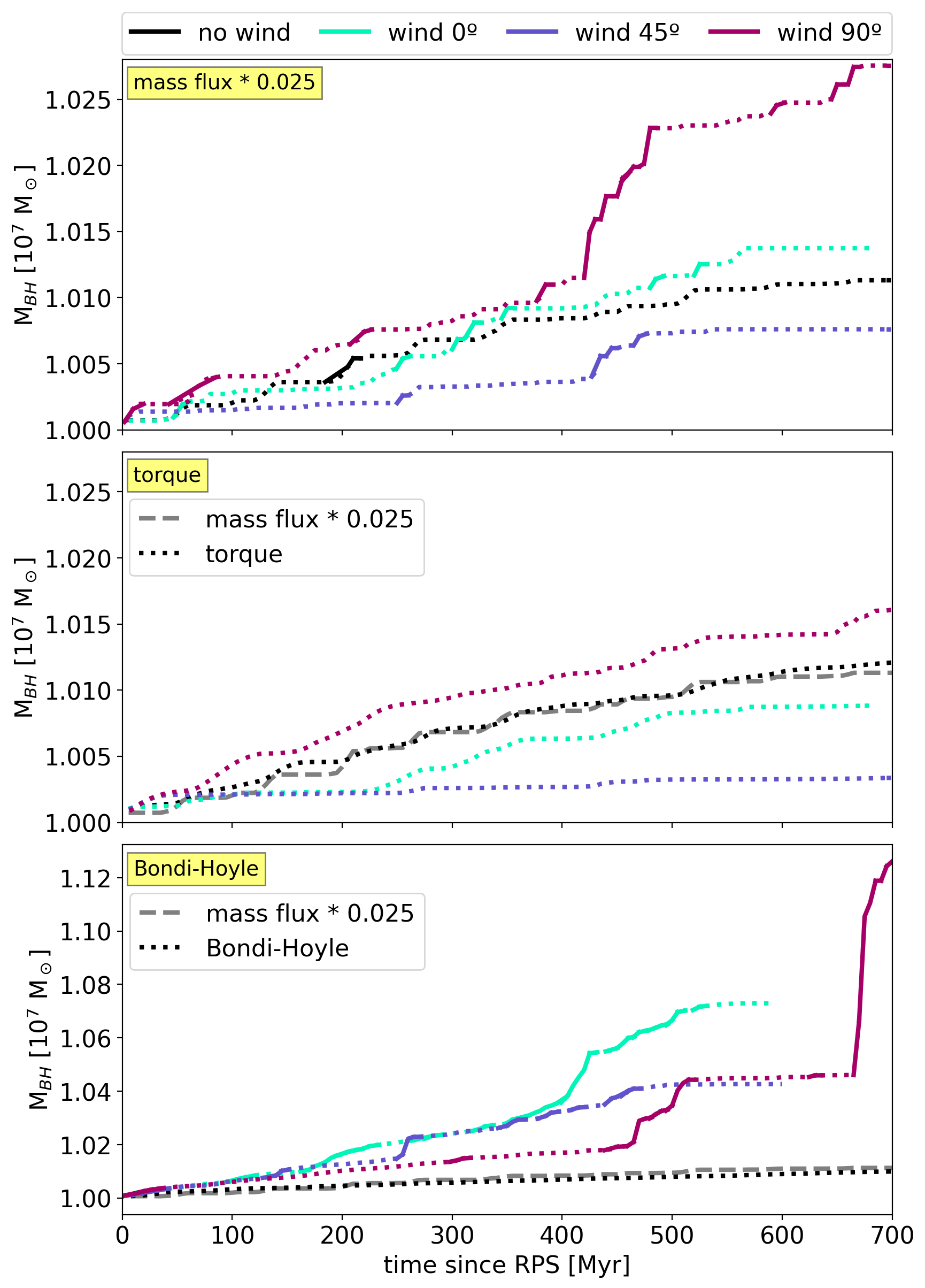}
\caption{BH mass as a function of time, measured by different accretion models and colour-coded by the wind angle. The dotted line represents regular BH growth, while the solid line indicates the AGN regime ($\ge$1 per cent of the Eddington accretion limit, see the text). In the middle and bottom panels, the dashed grey line shows the mass flux model for NW for comparison. Note that the Bondi-Hoyle accretion rate requires gas in a thin cylindrical shell at 700 pc from the galaxy centre, so can only be calculated while gas survives at this radius. This results in Bondi-Hoyle accretion ending earlier in W0 and W45 than when using the torque accretion model. In the top and middle panels y-axis range is the same, while the bottom panel has a bigger range.}
\label{Fig:BH_acc}
\end{figure}

Figure \ref{Fig:BH_acc} plots BH mass estimated using mass fluxes, the torque model and Bondi-Hoyle (panels from top to bottom), and we show the accretion rates themselves in Appendix \ref{appendix:Eddington}. The dotted line shows regular BH growth, while the solid line indicates the AGN regime, which for this work we define as a BH accretion rate at or above 1 per cent of the Eddington accretion limit for a $10^7 M_\odot$ BH. Although AGN have a range of Eddington ratios \citep{Trump09, Trump11, BestHeckman12}, this limit can guide our intuition for whether RPS is likely to boost BH accretion towards AGN luminosities and is motivated by the literature \citep{Heckman04, Ho09, KauffmannHeckman09}.

While the four simulations are shown as different coloured lines in each panel, in the middle and bottom panels we show the mass flux model for the NW galaxy as dashed grey line, for comparison. All three models predict similar BH growth for the NW galaxy. This is by design, as we chose $R_\text{B-H}$ and $\beta$ so that accretion rates from different models would match for the fiducial (NW) galaxy.

On the other hand, in the RPS galaxies the BH growth predictions are distinctly different for each of the estimators. We first focus on the mass flux estimator in the top panel. We have seen how under RP gas starts inflowing to the galaxy centre at an increased rate, even as net flux (Figure \ref{Fig:mass_flux}). This trend continues at lower scales (Figure \ref{Fig:BH_acc} top), with W90 accreting the most gas, followed by W0, NW, and with W45 accreting the least gas. We note that this also agrees with the relative masses in the central 500 pc in Figure \ref{Fig:mass_500}. Meanwhile, under RP, galaxies spend more time in the AGN state, with W90 being the most `active' one.

We next move to the middle panel to examine the torque model in detail. Although the relative predicted BH accretion rates among the RPS galaxies are the same as with the mass flux prediction, with W90, W0, and W45 accreting the most to the least mass, the actual accreted masses are much less than using a simple mass flux scaling. In fact, using the torque estimator W0 accretes less mass to a central BH than the NW galaxy, and W45 shows almost no growth at all, and none of the galaxies go through an AGN stage.

If we look at the variables in eq. \ref{eq:BH_acc}, it turns out that the time evolution of BH accretion rate is almost exclusively driven by the varying gas mass within $R_0$. This is because, in our simulations, the bulk of stellar mass is represented by static potentials and is therefore constant (Section \ref{sec:methods}). Figure \ref{Fig:dens_proj_centre} visually confirms that there is less gas in the centre of W0 and W45 compared to NW and W90.

To check the dependence on the central radius used to determine the variables for the torque model, we measured accretion using different $R_0$: 100 and 200 pc. Qualitatively it does not change the result. The only difference is that the higher $R_0$, the higher the accretion peak at the end of W90. Regardless of $R_0$, W90 accretes more gas than the NW galaxy, W0 accretes a bit less and W45 much less than the fiducial model.

By contrast, the Bondi-Hoyle model (Figure \ref{Fig:BH_acc} bottom) predicts that all three RPS-galaxies go through a lengthy AGN stage. Unlike the previous two models, the W0 and W45 galaxies have more early BH growth than the W90 galaxy. Only at late times, after W0 and W45 are completely stripped, does the W90 BH growth suddenly spike. We note that the BH growth in the wind runs reflects the increased mass fluxes in Figure \ref{Fig:mass_flux} (bottom), in particular, for W0 from 100 to 450 Myr, for W45 from 200 to 500 Myr and near the 400 and 600 Myr time steps for W90.

Unlike with the torque model, the speed of sound drives the time evolution of the Bondi-Hoyle accretion rate. This is due to the increasing clumpiness and growing holes in the ISM in Figure \ref{Fig:dens_proj_centre}. We see the increasing amount of ICM in galaxy centres of the RPS galaxies; this gas being hot drives up the average speed of sound. While increasing, the sound speed is still lower than the angular velocity for all RPS galaxies until 500 Myr, resulting in very little change of the denominator of equation \ref{eq:Bondi} while increasing the numerator.

We also note that the gas surrounding the BH ($R\leq140$ pc) has a mean density of $10^{-23}$ g cm$^{-3}$. The amount of mass in the central 140 pc would not be significantly changed in any of our BH accretion models, therefore we expect the average density would also be unaffected. Hence, our assumption that we can read the mass and density within 140 pc directly from the simulation without attempting to make corrections for mass added to the BH would not alter our results for either the Bondi-Hoyle or torque estimators.

To summarise, the three tested BH accretion models predict very different accretion rates for RPS-galaxies, despite all agreeing on the accretion rate in the fiducial NW run. The mass flux and Bondi-Hoyle models agree that RPS galaxies spend more time as AGN compared to NW, while in the torque model no galaxies reach the AGN state.

The models do all agree that W90 accretes the most gas and shows late jumps in its BH growth, possibly due to the inflow of dense gas seen in Figures \ref{Fig:dens_proj_centre} and \ref{Fig:mass_500}. We note that the majority of the W90 BH growth in the Bondi-Hoyle model comes after all the gas is stripped in W0 and W45, before which time BH growth is low, likely due to the lack of central hot ICM holes until late times. 

\section{Discussion} \label{sec:discussion}

\subsection{Comparison with JO201}

As we base our simulations on galaxy JO201, we also include in Figure \ref{Fig:tot_mass} the approximate gas mass of the galaxy \citep{GASPII} and where it should be located based on its position and velocity and on the density of its parent cluster \citep{GASPIX, ErratumGASPIX}. Since the wind angle of JO201 is $\approx40^\circ$ \citep{GASPXXIX}, it is closest to our W45 galaxy, which, unlike JO201, gets completely stripped by 600 Myr. One possible explanation is that the gas to stellar mass ratio might have been higher in JO201 prior to its falling into the cluster. Furthermore, the orbital parameters may differ between the simulations and observations, for example we could have overestimated RP values or constant wind angles could be an oversimplification that could change the stripping rate if a galaxy started edge-on, like W90. \cite{Vulcani18} estimated SFR in JO201 as $6\pm1M_\odot \text{yr}^{-1}$. In our simulation W90 has similar SFR, even though, again, the estimated wind angle of JO201 is closer to that of W45 that gets quenched to $<2M_\odot \text{yr}^{-1}$. However, we stress that, our goal in this work is not to simulate JO201, but to simply select initial conditions based on observational data.

% \vfill\null 

\subsection{Comparison with previous simulations}

\cite{Roediger06} model constant RP, but they find that the dependence of stripping rates on the wind angle is similar to ours. Namely, that galaxies with wind angles $< 60^\circ$ evolve similarly to a face-on stripped galaxy. They also note that the tail does not always point in the direction opposite of galaxy movement. We briefly mention this in Figure \ref{Fig:dens_proj_30}, as the W90 tail points to the right, as a result of the combined effect of RP and galaxy rotation. This should be taken into account when analysing observational data.

One of the first to hint at the possibility of increased inflows to the galactic centre were \cite{SchulzStruck01} who describe the gas at small galactic radii losing angular momentum and compressing. Similarly, \cite{Tonnesen09} find that ICM enters low-density regions and starts mixing with the rotating ISM, a process that we describe in Section \ref{subsec:mixing}, resulting in ISM angular momentum loss and inflow.

We have shown that the global SFR can be enhanced or quenched depending on the angle at which the ICM hits the galaxy (Figure \ref{Fig:SFH}). We expect that RP compresses the gas, which according to our SF prescription, leads to increased SFR. \cite{Bekki14} finds that the fate of the SFR under RP depends on several factors: galaxy mass, cluster mass and wind angle. SF enhancement is most probable for massive galaxies in low-mass clusters. We based our RP values (Figure \ref{Fig:tot_mass}) on JO201 falling into its parent cluster, Abell 85, that is a massive cluster with $M_\text{200} \simeq 10^{15} M_\odot$ \citep{GASPII}. Unlike here, in \cite{Bekki14} simulations SF is enhanced only temporarily and only at the pericentre passage for inclined galaxies \citep[see also][]{Steinhauser16}. What our simulations do agree on is that even in the best-case scenario (such as our W90) SF increase is not of star-burst type if we define it as an order of magnitude increase in SFR.

\cite{RuggieroLimaNeto17} model the effect of RP on a Milky-Way like galaxy as it falls into different clusters (with and without a cool core, with masses $10^{14}$ and $10^{15} M_\odot$). They describe a scenario similar to ours, in which SFRs increase immediately after a galaxy starts falling into a cluster (max 2 times enhancement) and quench during the pericentre passage if a galaxy gets completely stripped. They also show that SFR enhancement is higher under higher resolution as it allows to better capture gas compression.

\subsection{What causes the differences between mass flux and BH accretion estimators?} \label{subsec:discus_BH_acc}

In Section \ref{sec:BH_acc} we estimated BH accretion using three models: torque by \cite{HopkinsQuataert11}, Bondi-Hoyle modified by \cite{Tremmel17} and by mass flux \citep[normalised based on the relation reported in][]{Angles-Alcazar21}. Although we set these models to give similar accretion rates for the NW galaxy, they estimate totally different accretion rates for the galaxies undergoing RPS. Namely, the torque model shows that RPS-galaxies evolve closely to the NW one, that is, RP does not induce additional inflow to the BH and no galaxies reach the AGN state ($\ge$1\ per cent of Eddington accretion rate). Mass fluxes give similar results, only somewhat higher accretion for W0 and W45 than the torque model. W90 shows a dramatic late-time increase in the accretion rate that is only slightly hinted at in the torque model. The mass flux and Bondi-Hoyle models agree that RPS galaxies spend more time as AGN compared to NW, the Bondi-Hoyle model especially showing prolonged ($\approx100$ Myr) AGN stages. Here, we discuss the differences between the models and why they give such contrasting results.

We first begin by discussing the torque model. We remind the reader that the torque model is based on the assumption that stellar torques dominate the angular momentum removal of central gas \citep{HopkinsQuataert11}. This is not the case in our simulations, as we show in Section \ref{subsec:torques}. Torques in our fiducial model are dominated by pressure gradients, with much smaller gravitational torques from stars and gas.

This picture is different from \cite{Angles-Alcazar21} where stellar torques dominate (their Figure 16). One reason for this difference is that we use a static stellar potential, while in \cite{Angles-Alcazar21} all stars are represented by particles. Star particles that form in our simulation by 700 Myr account only for 2 per cent of the stellar disk mass in NW. As a result, our galaxies are more stable, and star particles follow a more homogeneous distribution (the higher the distribution asymmetry, the higher the torques). Moreover, our dark matter torques are zero because we include dark matter as a spherically-symmetric potential. However, we do note that \cite{Angles-Alcazar21} model a high-redshift galaxy ($2.28 < z < 1.10$) that is going through the stage of formation, so it is expected to be less stable than our galaxy that is based on JO201, which is fully-formed by $z=0.056$ \citep{GASPII}.

Although stellar torques may dominate due to a high stellar-to-gas mass ratio, we predict that gas torques should be more strongly effected than stellar torques by an angled RPS wind. As we discuss in Section \ref{subsec:torques}, the absolute value of the gas torque increases when dense arms of gas move near and through the galaxy centre (as seen in Figure \ref{Fig:dens_proj_centre}), which are formed by the wind pushing the ISM within the disk plane.

We argue that, even in a system in which stellar torques dominate, RP would drive pressure torques to increase more quickly than the stellar ones (Figure \ref{Fig:torque_cumu}), likely making them dominate or at least be of comparable value. Thus, we expect that our conclusion that the growing influence of torques from pressure gradients is what makes the torque model underestimate BH accretion (as we saw in Figure \ref{Fig:BH_acc}) is robust to galaxies with less symmetric stellar mass distributions.

While the torque model describes the accretion of cold, angular momentum supported gas, and the Bondi-Hoyle model describes spherically-symmetrical accretion of hot, pressure supported homogeneous gas, we see in Figure \ref{Fig:dens_proj_centre} that the distribution of these phases is asymmetric and differs between simulations. Therefore, because the two models outline different accretion conditions and processes, we would expect them to produce different accretion estimates.

The Bondi-Hoyle accretion rate increases with added mixing between the ICM and ISM and with the formation of large holes. These holes could create local pressure gradients either directly driving gas towards the centre or causing collisions that will drain the gas of orbital energy. In this respect, Bondi-Hoyle might actually be a better predictor for BH growth in RPS galaxies. However, the model has no way to account for the asymmetric pressure gradients and mass distributions from angled winds, so we do not anticipate that it will accurately predict the BH growth rate as a function of wind inclination angle.

Finally, the normalised mass flux measurement can indicate accretion trends, although we hesitate to trust the specific values, as we remind the reader that at 140 pc we are testing the limits of our resolution. Furthermore, the idea that net flux at a certain distance can represent BH accretion with a \textit{constant} factor might not hold for RPS-galaxies. \cite{Angles-Alcazar21} show (their Figure 10) that during the quasar phase of their AGN, the galaxy becomes more efficient at transporting the gas to the BH, as $\dot{M}_\text{0.1pc}/\dot{M}_\text{100pc}$ jumps by a factor of $\approx 25$ from $0.02$ in the other two phases. Of course, we do not expect RPS-galaxies to turn into quasars, but this illustrates that our assumption about the universal $\beta = 0.025$ is a simplification.

\subsection{Caveats}

\subsubsection{Resolution}

Our disk is resolved up to 39 pc, which as we show in Appendix \ref{appendix:resolution} is enough to resolve the movement of gas at 500 pc scales. It is when we estimate the BH accretion rate that we start to reach the limits of our resolution, as there are only 3 cells across a 140 pc radius. We also address this for the NW run in Appendix \ref{appendix:BH_resolution}, and find that a constant factor relates the flux at 500 pc and 140 pc, giving us some support for measuring BH accretion at 140 pc. RPS-galaxies require a more careful approach as shown in Section \ref{subsec:discus_BH_acc}, which makes it more difficult to correctly assess the accuracy of our BH accretion estimates. However, we note that our resolution is higher than in other simulations studying central inflows in RPS galaxies \citep{Ramos-Martinez18, Ricarte20}, and in future work we can more thoroughly test the impact of resolution using hyper-refined simulations as in \cite{Angles-Alcazar21}.

\subsubsection{BH sink particle}

As we have already noted, our simulation does not include a BH sink particle, so we can neither remove the accreted gas from the galaxy, nor exert AGN feedback. Not accreting the gas should not make a big difference, since as Figure \ref{Fig:BH_acc} and \cite{Angles-Alcazar21} suggest, gas consumption is a couple of per cent ($\beta = 0.025$ in eq. \ref{eq:flux_acc}) of the total net flux. \cite{Ricarte20} compare two similar galaxies subject to similar RP, of which one has a BH seed and the other does not. The galaxy with a BH briefly becomes an AGN after the pericentre passage. During this time, the AGN-driven outflows heat the surrounding gas and disturb the disk morphology, which as the authors note, might lead to rapid SF quenching. Of course, with the absence of a BH, no outflows develop in the other galaxy (and hence, in our simulations). In our future simulations, we plan to include a BH particle with AGN feedback.

\subsubsection{Missing physics}

Our simulations solve equations of hydrodynamics and include radiative cooling, star formation and stellar feedback, but still, we are missing some relevant physical processes.

For example, we do not include magnetic fields. \cite{Ramos-Martinez18} simulate a magnetised galactic disk subject to RPS by a non-magnetised ICM. They report that including magnetic field makes the disk flared, so that it is harder to strip the galaxy. In the initial conditions of our galaxy, the disk height decreases with radius, but as we turn on radiative cooling, the gas collapses into a disk with a radially-constant scale height. What is relevant to our work is that when the wind hits the flared magnetised disk, oblique shocks are produced, resulting in the flow of gas from the outskirts to the centre of the galaxy. This could further support our claim that RP induces inflows to the galactic centre, which might reach the BH and increase accretion compared to non-RPS galaxies. We note that we expect this process to be more active in face-on wind directions, like W0.

Furthermore, we do not include cosmic rays (CRs). \cite{Farber22} include both magneto-hydrodynamics and CRs and find that CRs suppress the inflow of gas to the galaxy centre. Nevertheless, similar to us, they find that under RP, these inflows are increased compared to the isolated galaxy even when accounting for CRs. 

Finally, although we do include SF and feedback, we have not experimented with other implementations. We are optimistic that our main conclusions will not be affected by a different SF recipe, as \cite{Lee20} use a different SF recipe with star formation efficiency that depends on the physical properties of the ISM and find that SF in enhanced in edge-on winds and depressed in face-on winds, in agreement with our simulations. Moreover, strong stellar feedback has been shown to restrict BH growth, particularly in low-mass $M_\text{star} < 10^9 M_\odot$ galaxies \citep{Habouzit17, Bower17}. We anticipate that a weaker feedback model would lead to a more dynamically cold ISM, possibly promoting stronger inwards gas motions, and thus reinforcing our conclusions. We note that experimenting with SF and feedback prescriptions in RPS simulations could be a useful avenue for constraining subgrid implementations of these processes.

\subsubsection{Model simplifications}

Our galaxy model includes stellar and gaseous disks and a DM halo, but does not include a bulge, even though this structure may be important for BH evolution as illustrated by tight correlations between bulge properties (velocity dispersion, luminosity and mass) and the BH mass \citep{Magorrian1998, Gebhardt00, Ferrarese&Merritt00, McConnell&Ma13, Batiste17}. However, some galaxies with psuedo-bulges instead of classical bulges show the same scaling relations \citep{Marasco21}, and some theoretical models even suggest that the correlation between the BH mass and bulge mass is non-causal \citep{King&Nealon21}. We choose to leave out the bulge in this simulation due to a lack of measurements for it in JO201, and the fact that the rotation curve from \cite{GASPII} does not indicate a significant central mass concentration. However, we expect that including a bulge would reinforce the conclusions made in this work, since bulges may drive the inflow of gas to the galaxy centre \citep{Park16}.

Another model simplification that we make is the exclusion of external tidal forces. As described in Section \ref{subsec:JO201}, our galaxy is approximately modelled after JO201 that is falling into its parent cluster (even though the model galaxy itself is fixed in space in the simulation box). We only assume the gravitational potential of the cluster itself to set the orbit parameters for the ICM wind. Realistically, a galaxy falling into a cluster could interact with other cluster galaxies or tides close to the cluster centre. Nevertheless, we purposefully neglect this mechanism to focus on the effects that pure RP has on gas in the centre of a galaxy. Moreover, as shown in \cite{GASPII}, JO201 is unlikely to have undergone tidal interactions, thus indicating that it is not a necessary condition for a galaxy to become an AGN.

\section{Conclusions} \label{sec:conclusions}

In this paper we run a suite of `wind-tunnel' galaxy-scale simulations to study the influence of a RP wind on the inflow of gas to the galaxy centre, focusing on the inner 500 pc to 140 pc. We base our galaxy model on a spectacular RPS galaxy JO201 which is slightly more massive than the Milky Way, and we compare an isolated galaxy not subject to RPS (no-wind, NW) to simulations modelling the galaxy on first infall into a cluster ICM, and vary the wind-galaxy impact angle: face-on (W0), edge-on (W90) and $45^\circ$ (W45). Our main results are the following:

\begin{enumerate}

\item Even though gas removal proceeds outside-in (Figure \ref{Fig:dens_proj_30}), in the central region of the disk visual differences in the ICM-ISM interaction appear as early as 100 Myr after the onset of RP (Figure \ref{Fig:dens_proj_centre}). Low-temperature holes rapidly develop in the centre of W0, but take longer to appear in W45 and W90.

\item The mass within the central 500 pc of the galaxy also differs in the NW and wind simulations, with both W0 and W90 clearly showing increased cold gas mass (Figure \ref{Fig:mass_500}). At all wind angles, RP increases both inflows and outflows of cold gas in the galaxy centre. Importantly, when we only consider the net flux, inflows under RP are higher than in the NW galaxy (Figure \ref{Fig:mass_flux}). 

\item In all RPS galaxies, ICM gas mixes with surviving disk gas in the central regions (Figure \ref{Fig:icmf_500}). We argue that this has a non-negligible effect on the amount of gas that could be in the galaxy centre, both due to accretion from the ICM, and due to angular momentum loss of the surviving ISM through mixing and the resulting inspiraling to lower orbits.

\item In addition, gas flows towards the galaxy centre due to torques mainly from local pressure gradients (Figure \ref{Fig:tot_torque}) that increase due to the pressure gradient along the wind direction. In the two most asymmetric galaxies, W90 and W45, pressure torques drive gas inflow, while in the most symmetric wind, W0, they tend to drive outflow.

 \item Under RP, standard BH accretion models do not predict similar accretion rates, neither quantitatively (Bondi-Hoyle predicts $10^2$ times more accretion than torque) nor qualitatively (until gas is completely stripped, Bondi-Hoyle predicts the most accretion in W0, while the torque model always has faster BH growth in the W90 galaxy).

\end{enumerate}

In Section \ref{subsec:discus_BH_acc} we connect the different predictions from the torque and Bondi-Hoyle accretion models to the RP-driven torques and mixing in the galaxies. We argue that the torque model underestimates accretion in RPS galaxies because it does not account for the increased torque due to pressure gradients driving inflow in W45 and W90. On the other hand, the Bondi-Hoyle model may overestimate BH accretion because of the hot ICM wind flowing through holes in the galaxy. We do note that because the hot ICM wind increases the speed of sound, some of the effects of RP may be indirectly included, such as gas mixing and large holes driving gas inflow - asymmetric effects cannot be part of the Bondi-Hoyle model, however.

For now, it is hard to answer the main question, that is whether RP turns galaxies into AGN. However, if we require an accretion rate of at least 1 per cent of Eddington limit for AGN, then RP increases the amount of time BH spend as AGN using either mass flux or Bondi-Hoyle (and the torque model simply never produces high accretion rates in any of our galaxies). As well as finding enhanced inflows to the galaxy centre, we identified some of the responsible mechanisms. In addition to important physical understanding, this will allow us to update BH feeding models to account for the processes that drive inflow in RPS galaxies.

In our next simulations we will add a BH seed, a sink particle which is able to accrete gas from the surroundings, and AGN feedback. As we have shown, we anticipate that this requires a new accretion model, and we could work on the basis of the torque model, but including a prescription for the increased local pressure torques. 

Bulges and bars are known to induce the inflow of gas to galaxy centres. We also know that galaxy JO201 that this simulation was based on has both the structures. Hence, adding bulge and bar potentials to the code will increase physical accuracy of the model and open up a discussion on the role of these structures in driving gas to a central BH in a galaxy undergoing RPS.

In short, we find that RP can affect the gas in the central region of a massive galaxy even at early times (and correspondingly low RP strengths). This effect is not gas removal, but accretion towards the centre, which we have shown could plausibly feed a BH.

\section*{Acknowledgements}

The authors would like to thank the referee for comments that improved the paper. We acknowledge the CINECA award under the ISCRA initiative, for the availability of high performance computing resources and support. This project has received funding from the European Research Council (ERC) under the European Union’s Horizon 2020 research and innovation programme (grant agreement No. 833824). The authors would like to thank Greg Bryan for help with the initial simulation set-up, and Daniel {Angl{\'e}s-Alc{\'a}zar} for useful discussions about black hole feeding models. We are also grateful to Benedetta Vulcani, Alessia Moretti, Marco Gullieuszik, Alessandro Ignesti and Andrea Kulier for their comments.

\bibliography{GASP}{}
\bibliographystyle{aasjournal}

\begin{appendix}
\restartappendixnumbering

\section{Resolution test} \label{appendix:resolution}

\begin{figure}[b]
\centering
\includegraphics[width = 85mm]{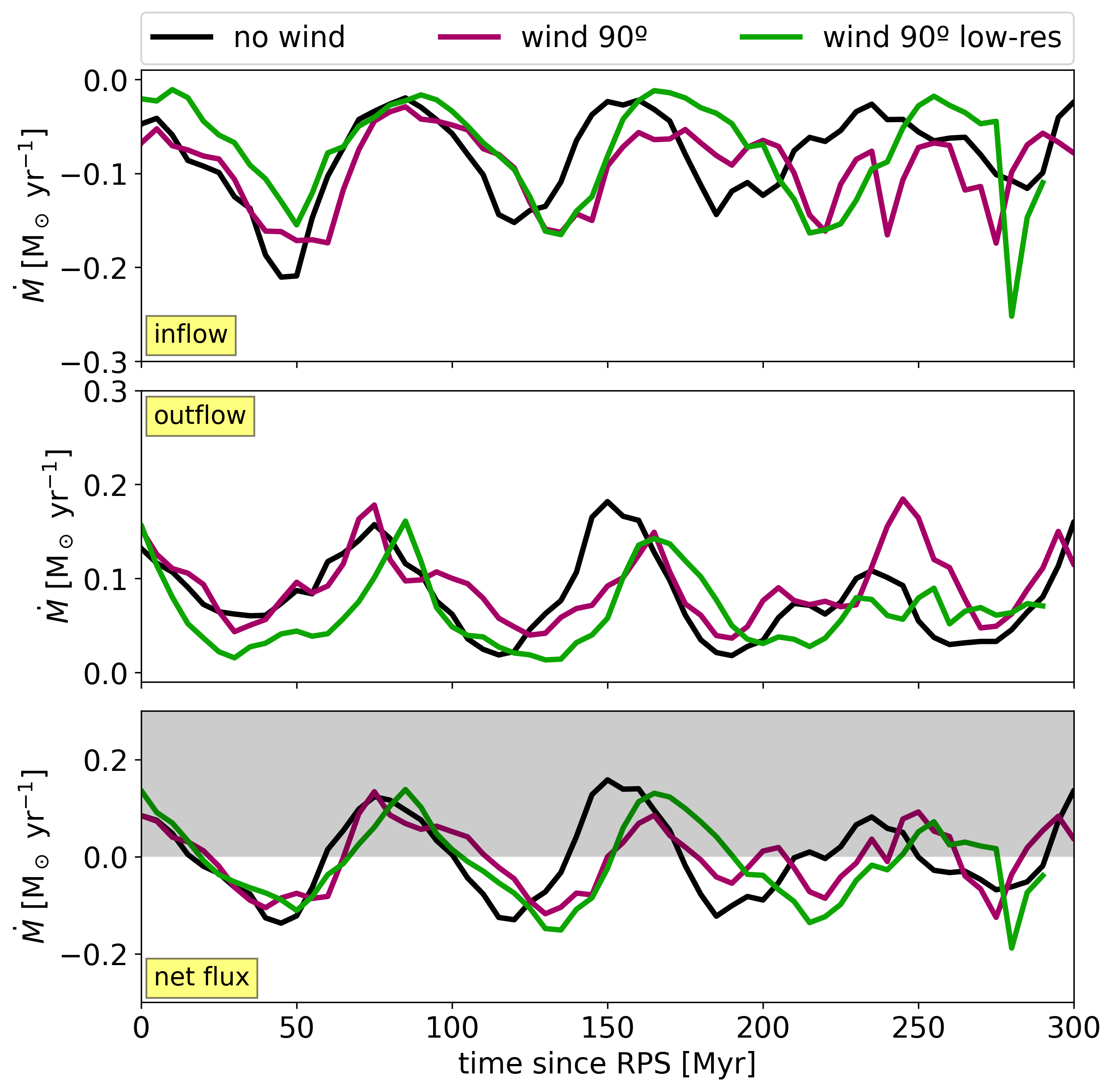}
\caption{Same as Figure \ref{Fig:mass_flux}, but colour-coded by simulation type.}
\label{Fig:mass_flux_low_res}
\end{figure}

\begin{figure}[b]
\centering
\includegraphics[width = 85mm]{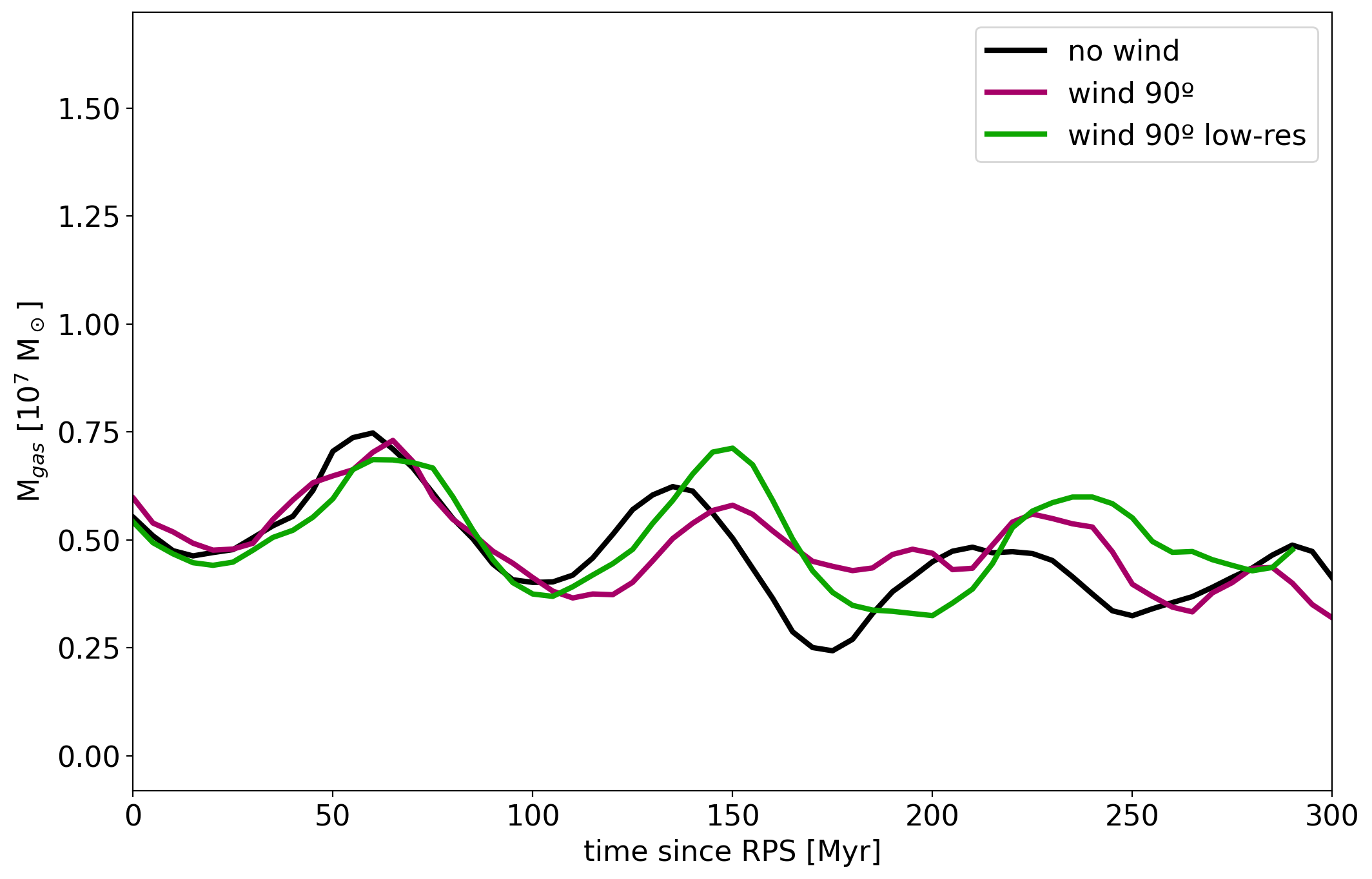}
\caption{Same as Figure \ref{Fig:mass_500}, but colour-coded by simulation type.}
\label{Fig:mass_low_res}
\end{figure}

To verify that our results are resolution-independent we run an additional short simulation of W90 galaxy. In the standard resolution runs we discuss throughout the paper, we include 5 levels of refinement allowing for the smallest cell size of 39 pc. Here, we decrease the number of refinement levels to 4, with a maximum resolution of 78 pc.

In Figure \ref{Fig:mass_flux_low_res} we repeat Figure \ref{Fig:mass_flux} comparing the fiducial and the low-resolution W90 runs. Here we show that the inflow, outflow and net mass flux are almost independent of resolution, probably because on 500 pc scales (shell width is 200 pc) the cold gas ($T < 10^{4.5}$ K) is well resolved even with 78 pc cells. Then, in Figure \ref{Fig:mass_low_res} we repeat Figure \ref{Fig:mass_500}, measuring cold gas mass within a sphere of $R=500$ pc. Again, we see that the mass in the galaxy centre is practically independent of resolution.

\section{Resolution test for BH accretion measurement} \label{appendix:BH_resolution}

\begin{figure}
\centering
\includegraphics[width = 85mm]{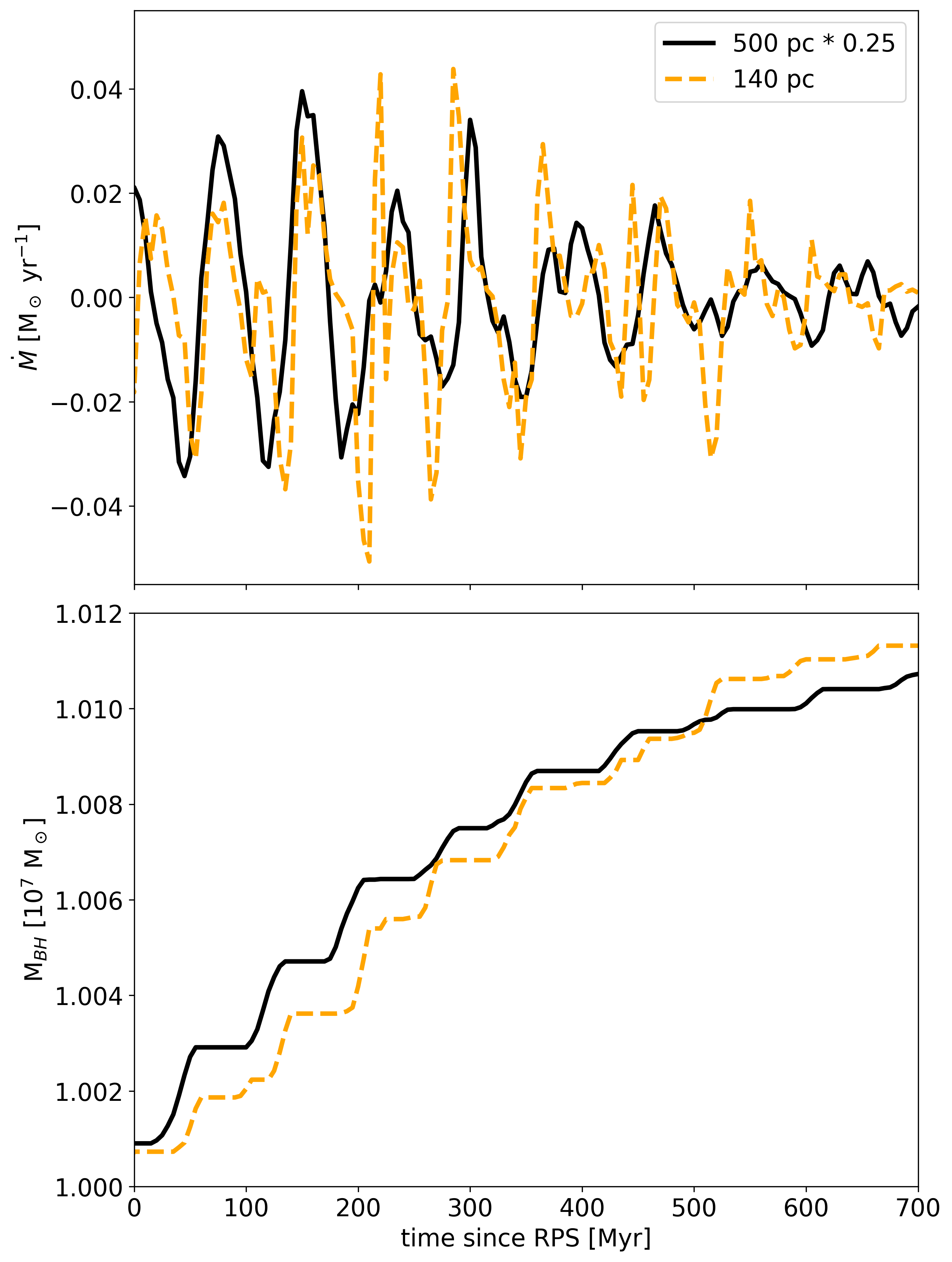}
\caption{Top: comparison between net flux measured in NW at 500 pc (eq. \ref{eq:mass_flux}) multiplied by a factor of 0.25 (solid black) and at 140 pc (Section \ref{subsec:BHflux}, dashed orange). Bottom: BH mass measured as in Section \ref{subsec:BHflux}, additional $\beta = 0.025$ factor is applied (eq. \ref{eq:flux_acc}).}
\label{Fig:res_test}
\end{figure}

To estimate the BH accretion rate with mass flux in Section \ref{subsec:BHflux} we measure cold gas mass flux across a shell 140 pc from the galaxy centre (80 pc width). With our standard 39 pc resolution, there are only 6 cells across a sphere of this radius, with 2 cells across the shell itself. To test how well we can resolve this region, let us compare mass flux in 140 pc with the one measured in 500 pc (eq. \ref{eq:mass_flux}) for the NW galaxy, under the assumption from Appendix \ref{appendix:resolution} that mass flux at 500 pc is relatively well-resolved.

In the top panel of Figure \ref{Fig:res_test} we show the two net fluxes, where we apply a constant factor of 0.25 to the 500 pc measurements with the same reasoning as in Section \ref{subsec:BHflux}. As we can see, the net flux at lower radius is very similar to the one in the outer shell, retaining the same periodical changes. In the bottom panel we plot BH mass estimated as in eq. \ref{eq:flux_acc} for the two mass fluxes (with the additional constant factor for the 500 pc). Here too, we see strong agreement between the two measurements, supporting our BH accretion estimation at 140 pc.

Note that we would only expect this 0.25 factor to hold in the NW run because as we have shown, a RPS wind affects the galaxy disk across all radii. 

\section{Late wind} \label{appendix:delayed}

\begin{figure}
\centering
\includegraphics[width = 85mm]{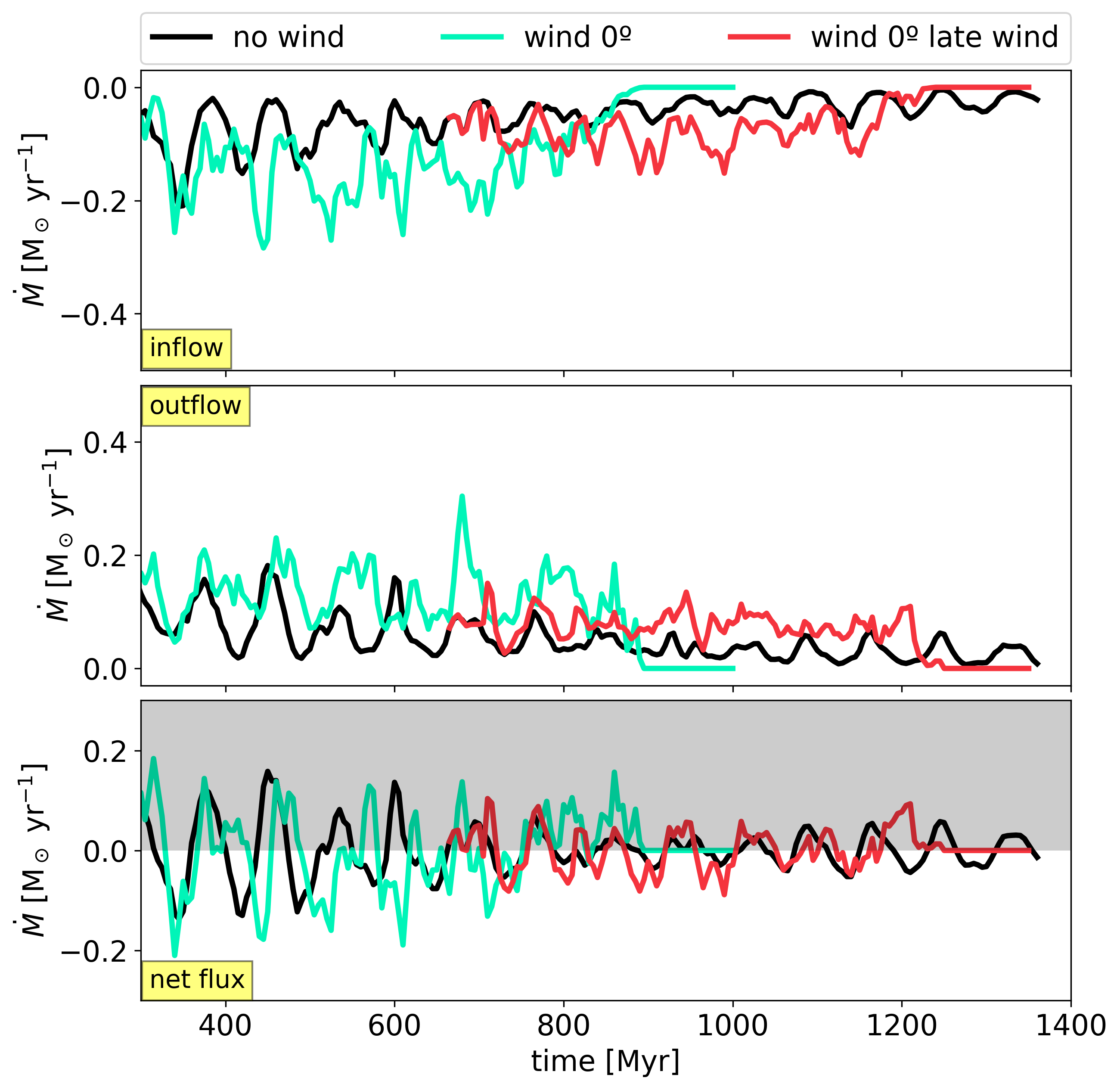}
\caption{Same as Figure \ref{Fig:mass_flux}, but colour-coded by simulation type. Note that the x-axis starts at the same time with the x-axis on Figure \ref{Fig:mass_flux} (300 Myr = 0 Myr since RPS).}
\label{Fig:mass_flux_late_wind}
\end{figure}

\begin{figure}
\centering
\includegraphics[width = 85mm]{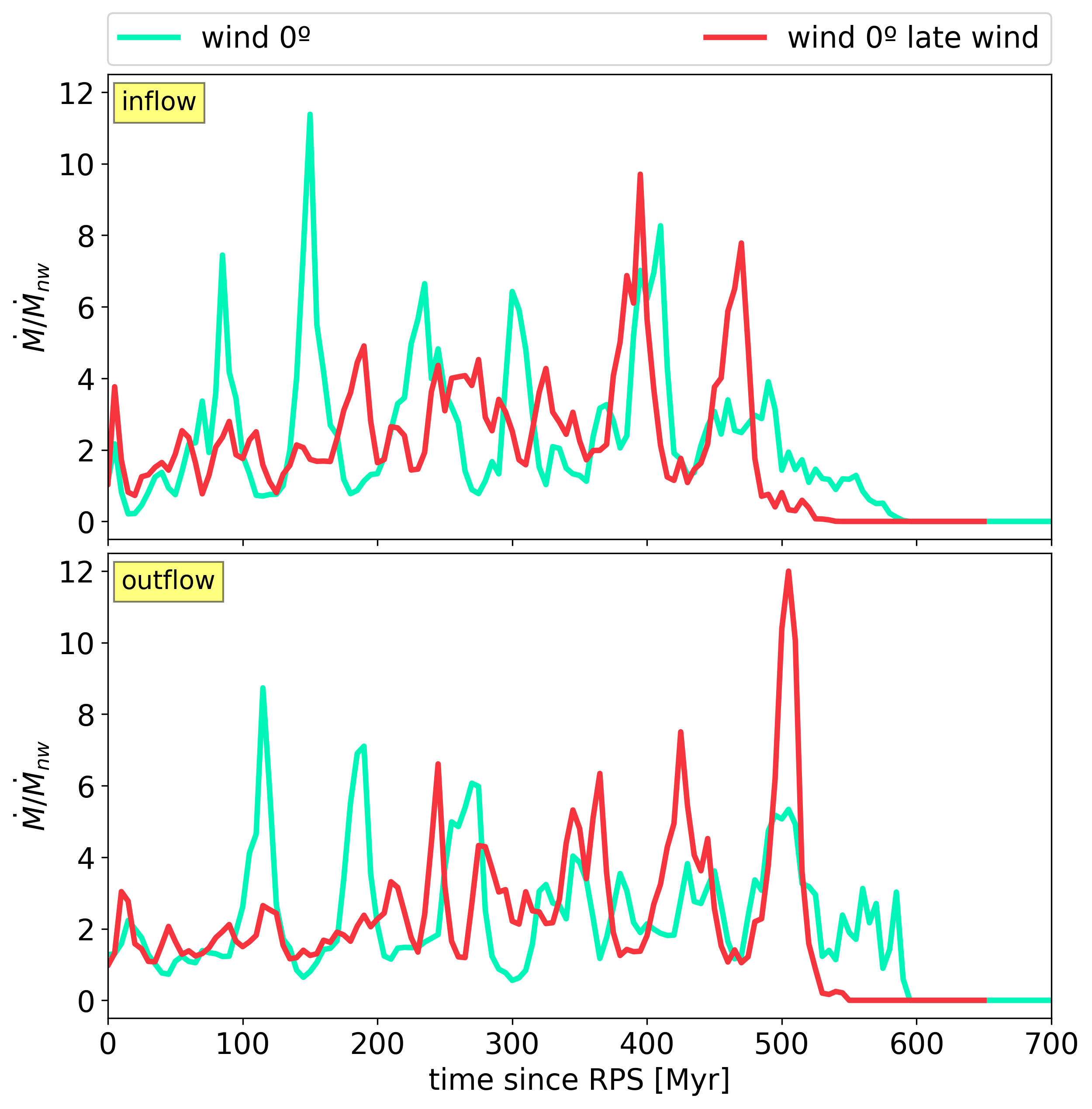}
\caption{Cold gas mass fluxes of RPS galaxies relative to NW mass fluxes, colour-coded by simulation type. \textit{Top}: relative inflows, \textit{bottom}: relative outflows. Mass flux is measured in a 200 pc wide spherical shell centred at 500 pc from the galaxy centre. Here, $t = 0$ marks the start of RPS, before which the galaxies were allowed to evolve for different times (Figure \ref{Fig:mass_flux_late_wind}).}
\label{Fig:mass_flux_late_wind_rel}
\end{figure}

As we discussed in Section \ref{sec:centre_gas}, the oscillations in mass and mass fluxes in the central region are mainly driven by the initial collapse of gas in the disk from radiative cooling and gradually decrease with time. In our fiducial runs, the galaxy evolves for 300 Myr before the wind hits it (Section \ref{sec:methods}). Here, to test that starting the wind at later times does not qualitatively affect our results we simulate a face-on stripped galaxy and start RP 400 Myr later, allowing for the total pre-wind evolution of 700 Myr.

In Figure \ref{Fig:mass_flux_late_wind} we again repeat Figure \ref{Fig:mass_flux}, only this time the x-axis is simply `time', where 300 Myr = 0 Myr since RP. Because in NW mass fluxes are decreasing with time, we need to compare the fiducial W0 and the late-wind W0 to the corresponding time steps in NW. Generally, we see that our main point stands -- under RP, both inflows and outflows are increased, and that in the net flux inflow peaks are higher.

In absolute values mass fluxes in late-wind W0 are lower than in the fiducial W0, again, because of the decreasing nature of fluxes in NW. To account for that, in Figure \ref{Fig:mass_flux_late_wind_rel} we plot mass fluxes in the two RPS-galaxies relative to mass fluxes in NW (at the corresponding times). Top panel shows relative inflows and bottom plots relative outflows, which illustrate that both fiducial and late-wind W0 evolve similarly. 

Thus we conclude that starting the wind at later times is unlikely to qualitatively affect our results.

\section{BH accretion rate per Eddington accretion rate} \label{appendix:Eddington}

\begin{figure}
\centering
\includegraphics[width = 85mm]{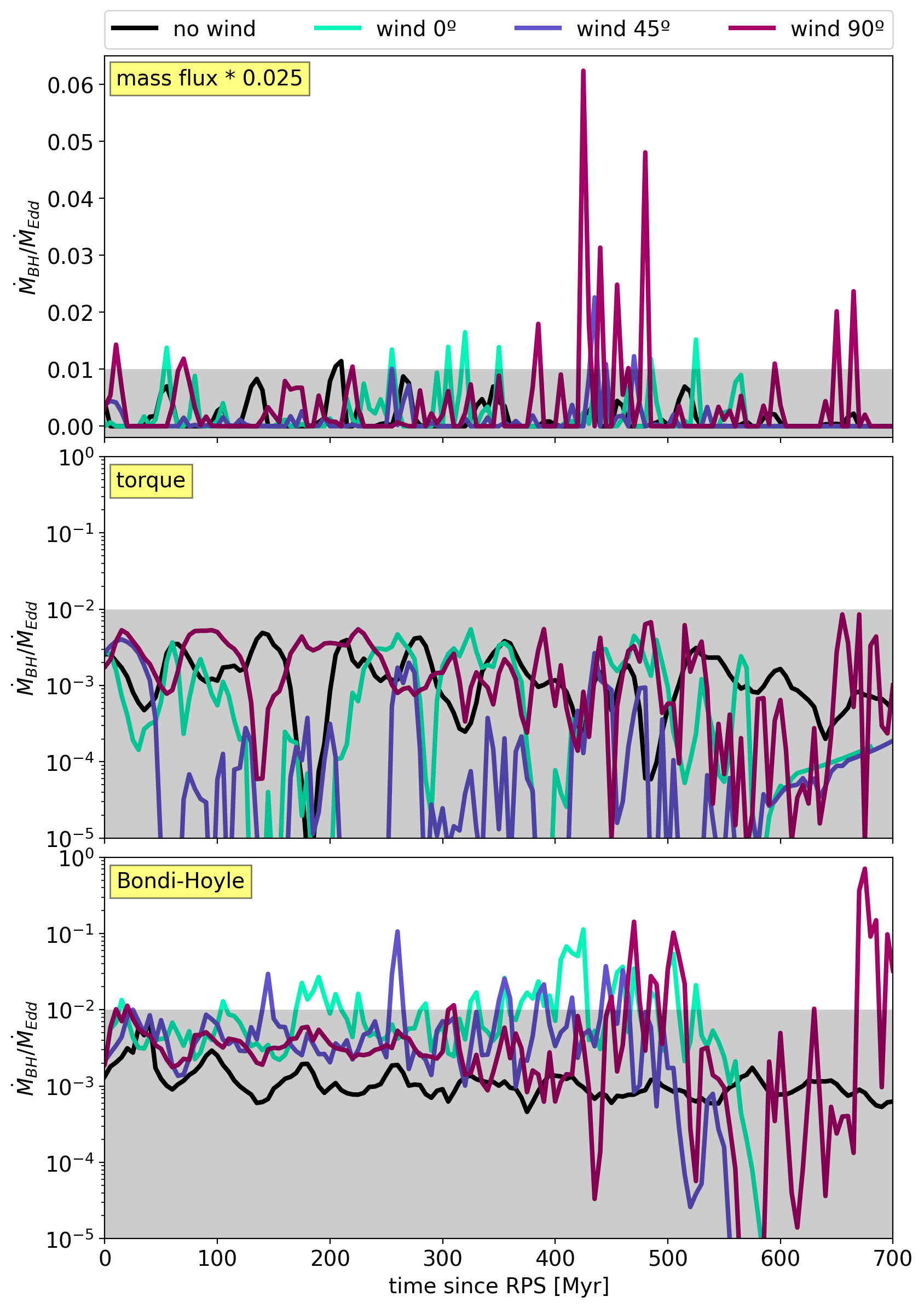}
\caption{BH accretion rate per Eddington accretion limit as a function of time, measured by different accretion models (Section \ref{sec:BH_acc}) and colour-coded by the wind angle. The grey area serves to separate the non-AGN regime from the AGN ($\ge$1 per cent of the Eddington accretion limit). In the middle and bottom panels y-axis has the same range and is in log scale, while the top panel has linear scale.}
\label{Fig:BH_Edd}
\end{figure}

To support our statements made in Section \ref{BH:comparison}, here, in Figure \ref{Fig:BH_Edd} instead of BH mass we plot BH accretion rates per Eddington accretion rate, calculated as $0.11 M_\odot \text{yr}^{-1}$ for a $10^7 M_\odot$ BH. We have posited that the AGN regime is 1 per cent of the Eddington accretion rate, and we illustrate this value by shading the non-AGN regime. Here, again, we see the distinctively different results provided by each of the accretion models for RPS-galaxies, not only quantitatively, but also in their relation with respect to the NW galaxy.

The torque model never predicts that any of the galaxies reach the AGN regime, while there are multiple short peaks in all wind runs above our AGN limit in both the mass flux and Bondi-Hoyle estimations.

\end{appendix}

\end{document}